\documentclass[twocolumn]{aastex631}
\usepackage{xcolor}
\usepackage{amsmath}
\usepackage{appendix}
\usepackage{soul}

\newcommand{\caiihk}{\ion{Ca}{2} H \& K}

\newcommand{\Msun}{\ifmmode {{M_\odot}}\else{$M_\odot$}\fi}
\newcommand{\Rsun}{\ifmmode {{R_\odot}}\else{$R_\odot$}\fi}
\newcommand{\Lsun}{\ifmmode {{L_\odot}}\else{$L_\odot$}\fi}

\shorttitle{Theia 456 Dynamical Analysis}
\shortauthors{Tregoning et al.}

\begin{document}

\title{Theia 456: Tidally Shredding an Open Cluster}

\correspondingauthor{Kyle Tregoning}
\email{kyle.tregoning.astro@gmail.com}

\author[0009-0009-6352-4964]{Kyle R.~Tregoning}
\affiliation{Department of Physics, University of Florida, 2001 Museum Rd, Gainesville, FL 32611, USA \\
}

\author[0000-0001-5261-3923]{Jeff J.~Andrews}
\affiliation{Department of Physics, University of Florida, 2001 Museum Rd, Gainesville, FL 32611, USA \\
}
\affiliation{Institute for Fundamental Theory, University of Florida, 2001 Museum Rd, Gainesville, FL 32611, USA }

\author[0000-0001-7077-3664]{Marcel A.~Ag\"{u}eros}
\affiliation{Department of Astronomy, Columbia University, 550 West 120th Street, New York, NY 10027, USA}
\affiliation{Laboratoire d’astrophysiquede Bordeaux, Univ. Bordeaux, CNRS, B18N, Allée Geoffroy Saint-Hilaire, 33615 Pessac, France}

\author[0000-0002-1617-8917]{Phillip A.~Cargile}
\affiliation{Center for Astrophysics | Harvard \& Smithsonian, 60 Garden Street, Cambridge, MA 02138, USA}

\author[0000-0003-2481-4546]{Julio Chanam{\'e}}
\affiliation{Instituto de Astrofísica, Pontificia Universidad Católicade Chile, Av. Vicuña Mackenna 4860, 782-0436 Macul, Santiago, Chile}

\author[0000-0002-2792-134X]{Jason L.~Curtis}
\affiliation{Department of Astronomy, Columbia University, 550 West 120th Street, New York, NY 10027, USA}

\author[0000-0001-7203-8014]{Simon C.~Schuler}
\affiliation{University of Tampa, Department of Physics and Astronomy, Tampa, FL 33606, USA}

\begin{abstract}

The application of clustering algorithms to the Gaia astrometric catalog has revolutionized our census of stellar populations in the Milky Way, including the discovery of many new, dispersed structures. We focus on one such structure, Theia 456 (COIN-Gaia-13), a loosely bound collection of $\sim 320$ stars spanning $\sim120$ pc that has previously been shown to exhibit kinematic, chemical, and gyrochronal coherency, indicating a common origin. We obtain follow-up radial velocities and supplement these with Gaia astrometry to perform an in-depth dynamical analysis of Theia 456. By integrating stellar orbits through a Milky Way potential, we find the currently dispersed structure coalesced into a small cluster in the past. Via Bayesian modeling, we derive a kinematic age of $245\pm3$ Myr (statistical), a half-mass radius of $9\pm2$ pc, and an initial one-dimensional velocity dispersion of $0.14\pm0.02$ km s$^{-1}$. Our results are entirely independent of model isochrones, details of stellar evolution, and internal cluster dynamics, and the statistical precision in our age derivation rivals that of the most precise age-dating techniques known today, though our imperfect knowledge of the Milky Way potential and simple spherical model for Theia 456 at birth add additional uncertainties. Using posterior predictive checking, we confirm these results are robust under reasonable variations to the Milky Way potential. Such low density structures that are disrupted by the Galactic tides before virializing may be ubiquitous, signifying that Theia 456 is a valuable benchmark for studying the dynamical history of stellar populations in the Milky Way.

\end{abstract}

\section{Introduction} \label{sec:intro}

The Gaia catalog, with over 1.8 billion sources in its third data release, has transformed the study of stars and stellar populations within our own Milky Way and its surrounding satellite galaxies \citep{Gaia_mission, Gaia_DR3_release, Babusiaux_2023}. Of these 1.8 billion sources, roughly 80\% have full five-dimensional (5D) astrometric data with unparalleled precision. In particular, this wealth of high quality data has aided in uncovering numerous coeval populations, such as open clusters \citep{Castro-Ginard_2022, Hao_2022}, moving groups and young associations \citep{Faherty_2018, Ujjwal_2020}, extended coronae and tails connected to known clusters \citep[][]{Meingast2021},
as well as linking newfound structures to well-studied open clusters and moving groups \citep{Gagne_2021}. Along with revealing new structures, Gaia data has been instrumental in understanding star cluster kinematics \citep{Kuhn_2019, Vasiliev_2021}, Milky Way substructure \citep{Castro-Ginard_2021} and star formation history \citep{Kerr_2021, Briceno-Morales_2023}, and refining censuses for open clusters and associations \citep{Cantat-Gaudin_2018, Luhman_2020}. 

Star clusters are extensively used as benchmarks to study stellar evolution \citep[e.g.,][]{Kalirai_2010}, reliant on the self-consistent age and metallicity of their members. The need for accurate and precise age dating methods has been discussed extensively in the literature, each with its own benefits and drawbacks \citep[for a review, see][]{Soderblom_2010}. Model dependent methods, like isochrone fitting, can reach uncertainties as large as $50\%$, and rely heavily on accurate de-reddening corrections. Astroseismology, which utilizes models of stellar pulsations, can provide ages accurate to $\sim 10\%$, but relies heavily on high-resolution photometry and spectroscopy \citep[see][for an example]{Li_2022}. \cite{Skumanich_1972} uncovered an empirical relation between rotation period, age, and stellar activity by comparing 
chromospheric \caiihk\ 
emission for stars in the Pleiades, Ursa Major, Hyades, and the Sun. This discovery laid the foundation for gyrochronology \citep{Barnes_2003}, which relies on measuring the rotation period of a star, $P_{\rm rot}$, which is now relatively easy to observe and can age-date single stars. However, a poor understanding of the details of spin-down \citep[e.g., the temporary phase of stalling;][]{Curtis_2020} makes it especially difficult to age-date lower mass K- and M-type stars. Dynamical age-dating, utilizing the basic assumption that stars were likely born when they were most compact in position space, is typically only applied to relatively young open clusters \citep[e.g.,][]{Jilinski_2008}. For these reasons, nearby common origin structures, where all of the techniques detailed above and more can be applicable, are the predominant sources for age-dated stars in our Milky Way.  

It is widely believed that the bulk majority of star formation occurs in these clustered environments through the gravitational collapse of cold molecular gas \citep{Lada_2003}. However, only a small fraction of these systems survive as bound star clusters \citep[e.g.,][]{Bressert_2010, Kruijssen_2012}. The majority of these stars are destined to disperse into the Milky Way field \citep{Goodwin_Bastian_2006, Coronado_2020}. The evolution of stars from their clustered birth-places to their eventual position within the Milky Way field is not yet fully understood, though significant steps have been made \citep[see][for a review]{Helmi_2020}. For example, \citet{Kamdar_2019} reproduce the age--velocity relation, surface-density profile, and metallicity distribution function in the Solar Neighborhood via dynamical modeling of clustered star formation. 

Ever increasing astrometric precision from Gaia has spawned the discovery of many exciting new stellar structures that may probe this transition from cluster to field star: stellar streams found within the Milky Way disk. Prior to the advent of Gaia, only stellar streams within the Milky Way halo were discussed extensively in the literature. 
Within the disk, disentangling kinematically coherent streams from foreground and background stars has been made possible via the combination of machine learning and clustering algorithms with highly precise astrometry from Gaia. \cite{Kounkel_2019} and \cite{Kounkel_2020} apply \texttt{HDBSCAN} \citep{Campello_2013, McInnes2017}, a hierarchical clustering algorithm shown to be highly effective at recovering open clusters in the Gaia dataset across varying density scales \citep{Hunt_2021}, to the Gaia DR2 catalog \citep{Gaia_DR2}. They identify many phase space over-densities in search of disperse stellar structures; their Theia catalog contains 8292 structures, often extending hundreds of parsecs in length with hundreds or thousands of stars, all within 3 kpc of the Sun. \cite{Hunt_2023} also apply HDBSCAN to the Gaia DR3 catalog to improve the open cluster census, finding 739 high-probability open clusters previously unmentioned in the literature. Numerous similar searches for open clusters in the Solar Neighborhood and beyond have been performed recently in the literature \citep[e.g.,][]{He_2022, Chi_2023}. The combination of clustering algorithms with the Gaia catalog has shown that the Milky Way is ripe with coeval stellar populations, many of them low density, allowing us to better study the interplay between cluster and galactic dynamics.

One such object from the \cite{Kounkel_2019} catalog, Theia 456, extends nearly $120$ pc in length and 20$^{\circ}$ across the sky in the Northern Hemisphere, a typical length for a `string' in the Kounkel \& Covey catalog. This structure, which was originally discovered by \cite{Cantat-Gaudin_2019} and named COIN-Gaia-13, was shown to be a kinematically, chemically, and gyrochronally coherent, thin-disk stream by \citet{Andrews_2022}. Additionally, \cite{Bai_2022} use the unsupervised clustering algorithm \texttt{pyUPMASK} \citep{pyUPMASK} to determine astrometric membership of the extended structure, and find that it is likely in the process of dissolution via differential Galactic rotation. Despite these follow-up studies, imprecise radial velocity measurements among the subset of members observed by additional instruments, $\sim 5\; \rm km~s^{-1}$ precision for 24 stars with Gaia DR2 radial velocities \citep{Gaia_DR2} and $\sim 6\; \rm km~s^{-1}$ precision for 69 stars with LAMOST radial velocities \citep{Cui_2012, Luo_2015}, precluded an in-depth dynamical analysis. As we describe in Section \ref{sec: quantitative dynamical history}, a more detailed dynamical analysis of Theia 456 requires an RV precision of $\sim 1~\rm km~s^{-1}$ or better. Its closeness to the Sun ($\sim 500 ~\rm pc$), coeval nature, and two lobed structure make Theia 456 ripe for additional ground-based spectroscopic observations across its entire extent. 

Having obtained spectroscopic follow-up observations, we re-visit Theia 456, taking a much more detailed look at the dynamics of the system. In Section \ref{sec:data} we present new radial velocities from MMT's Hectochelle, a precise multiobject echelle spectrograph, and discuss the quality cuts we make to obtain our dynamical sample. We utilize this sample to provide a qualitative dynamical history of Theia 456 by integrating orbits in a Milky Way potential in Section \ref{sec:2 lobes of Theia 456}, and quantitatively estimate Theia 456's age, birth scale, initial velocity dispersion, and initial phase space position via a Bayesian approach in Section \ref{sec: quantitative dynamical history}. In Section \ref{sec:discussion}, we comment on relevant assumptions made related to the Milky Way potential and Theia 456's self-gravity, and propose potential applications of stellar streams as probes of star forming regions. We provide some concluding thoughts in Section \ref{sec:conclusions}. 

\begin{figure}
    \begin{center}
    \includegraphics[width=1.0\columnwidth]{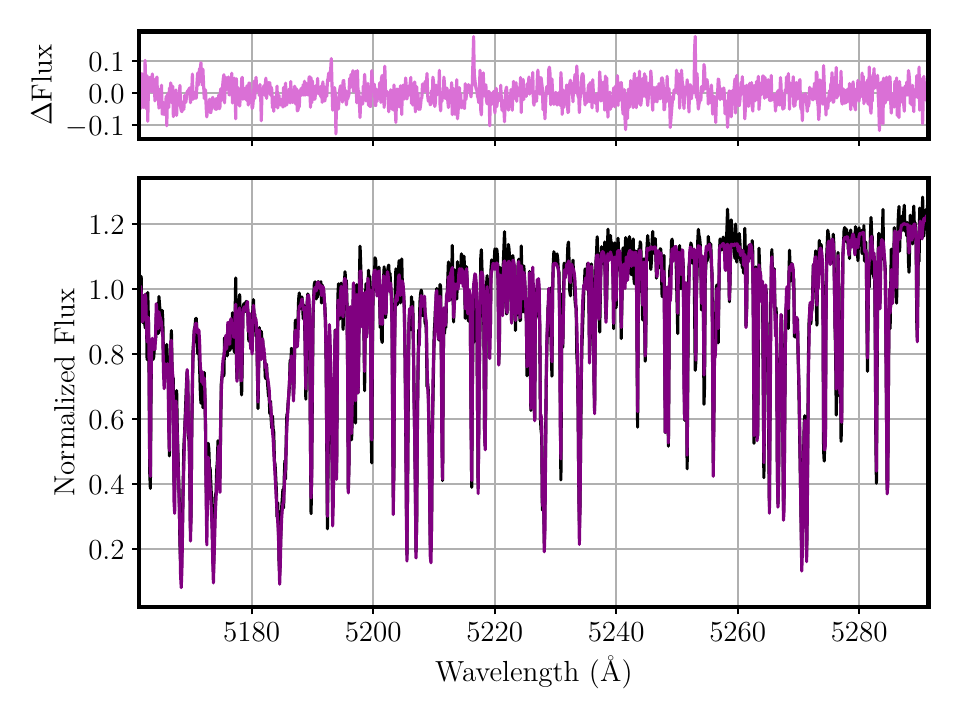}
    \caption{An example spectrum from MMT-Hectochelle for a member of Theia 456 (Gaia DR3 191357324369704448) in black, and the corresponding \texttt{MINESweeper} fit to the spectrum in purple. Residuals are shown in the top panel.}  
    \label{fig:example_spectrum}
    \end{center}
\end{figure}

\begin{figure}[h!]
    \begin{center}
    \includegraphics[width=1.0\columnwidth]{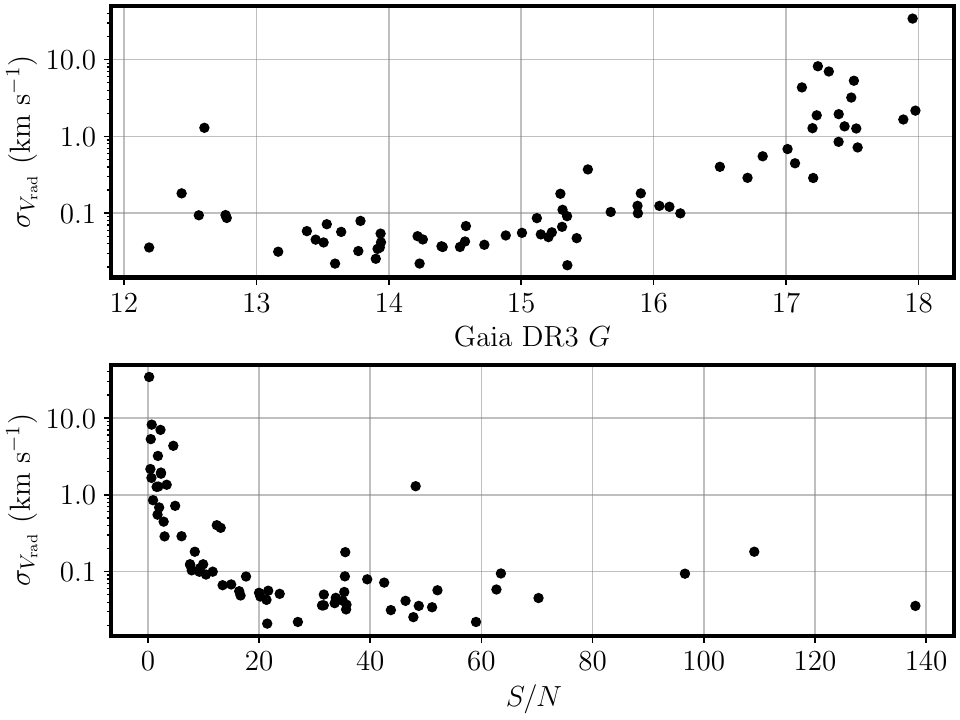}
    \caption{Measurement uncertainties for stellar radial velocities determined from Hectochelle spectra. \textit{Top}: Radial velocity precision plotted versus observed Gaia~DR3 $G$ magnitude. \textit{Bottom}: Similarly, radial velocity precision shown against the median signal-to-noise ratio of the Hectochelle spectrum.}  
    \label{fig:rverrors} 
    \end{center}
\end{figure}

\begin{figure*}
    \begin{center}
    \includegraphics[width=1.0\textwidth]{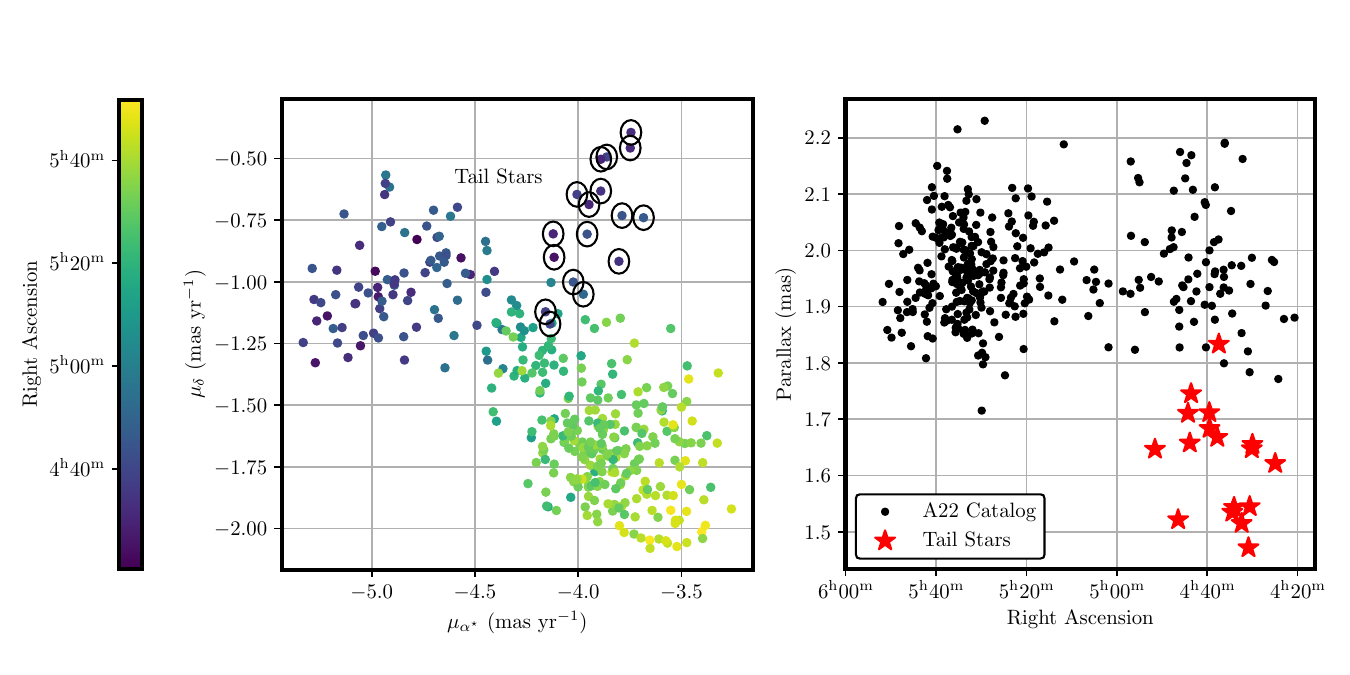}
    \caption{\textit{Left}: Proper motion space of Theia 456 stars from the A22 catalog, color-coded by right ascension, which is a proxy for position along the stream. Circled stars (labelled as tail stars) appear systematically offset from the observed positional gradient in proper motion space. \textit{Right}: Parallax vs. right ascension - tail stars have systematically lower parallaxes than other Theia 456 stars at the same right ascension. We cull these tail stars from our catalog.}  
    \label{fig:tail_stars}
    \end{center}
\end{figure*}

\section{Spectroscopic Data} \label{sec:data}

Data are obtained for members of Theia~456 with the Hectochelle multi-object spectrograph \citep{Szentgyorgyi2011} on the MMT. We use the RV31 filter setup resulting in spectra with a resolution of $R \sim 32,000$ over a single wavelength range of 5150--5300 $\rm \AA$. This instrument setup and wavelength range has been shown to provide reliable stellar parameters \citep[][]{Cargile2020}. The spectra were processed using the standard reduction pipeline for Hectochelle data (HSRED v2.1\footnote{\url{https://bitbucket.org/saotdc/hsred}}). 

We observe six fields along Theia~456, placing Hectochelle fibers on kinematically selected cluster members. Each field was carefully selected to simultaneously maximize the number of high-confidence, bright cluster members (down to $G$ $\sim$ 18 mag), while providing information about stars across its extent. We derive stellar parameters by modeling the spectra using the \texttt{MINESweeper} code \citep{Cargile2020}, which has been used extensively to model similar Hectochelle spectra \citep[e.g.,][]{Conroy2019}. Figure \ref{fig:example_spectrum} presents an example spectrum and its corresponding \texttt{MINESweeper} fit.

\subsection{Radial Velocities} \label{sec:radialvelocities}

Radial velocities are determined from spectral fitting for all Hectochelle data as output from the \texttt{MINESweeper} stellar parameter pipeline. Barycentric corrections were calculated using {\tt Astropy} \citep{astropy_v3} and applied for each observed MMT hectochelle observation. Our measured velocities have a median precision of $\sim$0.1 km s$^{-1}$, with precision only rising above $\sim$1 km s$^{-1}$ for the faintest, 
lowest signal-to-noise ratio spectra (i.e., $G$ $>$ 17, $S/N <$ 10 per pixel; see Figure \ref{fig:rverrors}).

\subsection{Our Sample} \label{sec:dynamical members}

In order to obtain a useful dynamical sample, we first refine the catalog of Theia 456 members reported in \cite{Andrews_2022} (hereafter A22) in order to accurately determine Theia 456's median astrometric quantities (i.e. $\mu_{\alpha^{\star}} \equiv \mu_{\alpha} \cos \delta$, $\mu_{\delta}$, $\varpi$) for comparison with the stars with Hectochelle radial velocities. We omit members that appear as a tail in proper motion space in the A22 catalog, with systematically larger values of $\mu_{\alpha^{\star}}$ compared with other stars within the same region of Theia 456, visible in Figure \ref{fig:tail_stars}. These stars also feature systematically lower parallaxes when compared with the rest of the stream; on average, these tail stars sit $0.3$ mas below the median stream parallax of $\varpi \approx 1.9 \pm 0.1 $ mas reported in A22. We cull these discrepant members from our catalog, reducing its size from 362 to 345 stars. 

\begin{figure*}
    \begin{center}
    \includegraphics[width=1.0\textwidth]{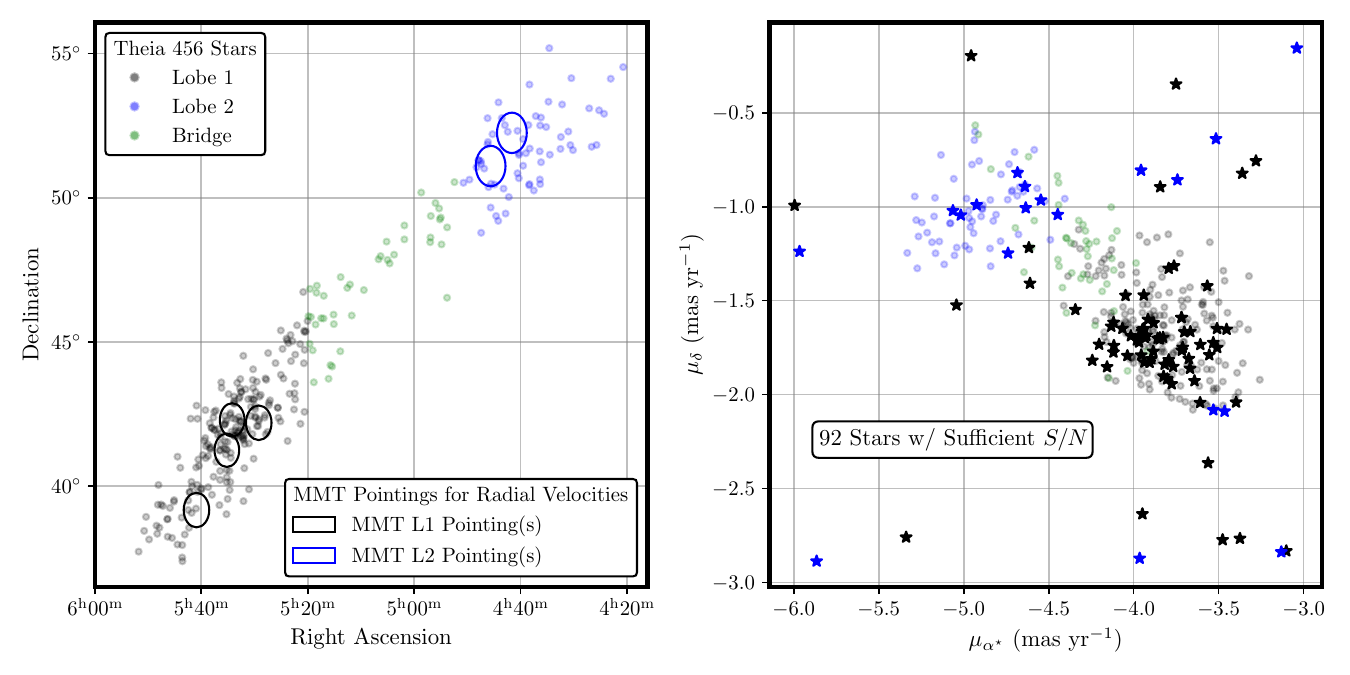}
    \caption{\textit{Left}: MMT telescope pointings used to obtain Hectochelle radial velocities, overlaid on our Theia 456 catalog. \textit{Right}: 92 stars from the Hectochelle sample with sufficiently precise radial velocities and metallicities and high signal-to-noise ratios, compared to all Theia 456 members in proper motion space. Circle markers denote stars from our catalog, while star markers denote stars from the Hectochelle sample with precise radial velocities. Black star markers come from the MMT pointings in the southern lobe (L1) of Theia 456, while blue star markers come from the northern lobe (L2).}  
    \label{fig:456_sample_andrews}
    \end{center}
\end{figure*}

\begin{figure*}
    \begin{center}
    \includegraphics[width=1.0\textwidth]{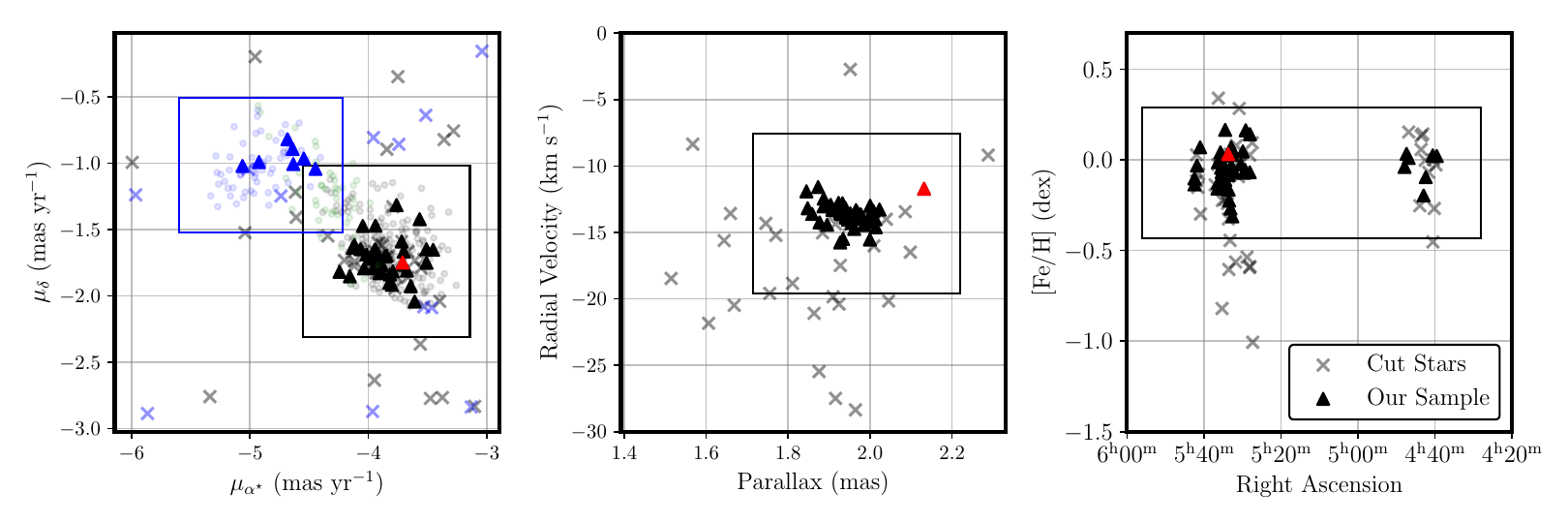}
    \caption{Plot of our sample and culled stars in various parameter spaces, with selection criteria shown as rectangles. Bold triangles show our sample, while crosses represent stars culled via our refinement process. In proper motion space (the left-most plot), we color-code stars and selection criteria by lobe; black for L1 and blue for L2, and overplot our sample on top of the entire Theia 456 catalog. Crosses that appear within the selection criteria were culled via some criterion not shown on that plot (i.e., crosses within the rectangle in radial velocity vs. parallax space were culled via proper motion or metallicity criteria). The red triangle shows a single star that passed all selection criteria, but was culled manually due to an inconsistent parallax and large parallax uncertainty ($\sigma_{\varpi} = 0.073$ mas, roughly three times larger than the median parallax uncertainty in our sample). } 
    \label{fig:kinematic cuts}
    \end{center}
\end{figure*}

We obtained a sample of 270 stars with highly precise radial velocities from MMT's Hectochelle spectrograph from six fields, shown in the left panel of Figure \ref{fig:456_sample_andrews}. These fields contain members of Theia 456, suspected members, potentially erroneous members, and background stars. We wish to define an additional catalog of stars containing both Gaia astrometric data and Hectochelle radial velocities; we undergo the following procedure to do so:

\begin{enumerate}
  \item Cull stars in the MMT-Hectochelle sample with large uncertainties in radial velocity ($>$ 5 km s$^{-1}$), $\mathrm{[Fe/H]}$ ($>$ 0.05 dex), or low signal-to-noise ratios ($<$ 5). This reduces our sample from 270 to 92 stars, visible in the right panel of Figure \ref{fig:456_sample_andrews}. 
  
  \item Separate stars in our catalog into two separate lobes in R.A.\ and decl.\ with a disperse bridge joining them. We designate stars with R.A. $> \mathrm{ 5^h 20^m}$ as one lobe (L1), and stars with R.A. $< \mathrm{ 4^h 52^m}$ as another (L2). Stars in between are labelled as bridge stars. This qualitatively matches the observed structure of Theia 456. The left panel of Figure \ref{fig:456_sample_andrews} shows that four MMT pointings fit firmly into L1, with the other two pointings in L2.  
  
  \item From our catalog of 345 stars with Gaia data, we compute the median and dispersion of $\mu_{\alpha^{\star}}$, $\mu_{\delta}$, and $\varpi$. We do this by lobe in proper motion space, and across the whole structure in parallax, because clear positional gradients exist in proper motion space, but not in parallax: 
  
  \begin{itemize}
    \item $\mu_{\alpha^{\star}}^{\rm L1}$: $-3.8 \pm 0.2$ mas yr$^{-1}$
    \item $\mu_{\alpha^{\star}}^{\rm L2}$: $-4.9 \pm 0.2$ mas yr$^{-1}$
    \item $\mu_{\delta}^{\rm L1}$: $-1.7 \pm 0.2$ mas yr$^{-1}$
    \item $\mu_{\delta}^{\rm L2}$: $-1.0 \pm 0.2$ mas yr$^{-1}$
    \item $\varpi$: $2.0 \pm 0.1$ mas
  \end{itemize}
  
  We restrict our sample of stars with Hectochelle radial velocities to stars with proper motions no further than $0.6$ mas yr$^{-1}$ ($\approx$ 1.5 km s$^{-1}$ at the distance of Theia 456) from the median quantity of its respective lobe, and within $0.3$ mas ($\approx$ 90 pc at the distance of Theia 456) of the median parallax. We show our astrometric selection criteria as boxes in Figure \ref{fig:kinematic cuts}; proper motion in the leftmost panel and parallax in the middle panel. 
  
  \item Restrict our sample to stars with metallicities no further than 0.36 dex away from the median ([Fe/H]: $-0.07 \pm 0.12$ dex, reported in A22 via analysis of LAMOST DR5 metallicities). This selection criterion is shown in the rightmost panel of Figure \ref{fig:kinematic cuts}.

  \item Compute the median Hectochelle-measured radial velocity of the remaining stars ($-13.6\; \rm km~s^{-1}$). At this point, many non-members have been filtered out, making the median radial velocity a reliable estimate of Theia 456's bulk radial velocity. We then cut stars further than $6\; \rm km~s^{-1}$ from the median, similar to the radial velocity dispersion found in Tian 1 \citep{Tian_2020}, a `stellar snake' with similar extent across the sky and distance as Theia 456. The middle panel of Figure \ref{fig:kinematic cuts} shows this selection criterion.
  
\end{enumerate}

Following this procedure, our dynamical sample contains 44 stars with Hectochelle radial velocities, from 270 stars observed. We cut one star from L1 manually, which has a particularly large uncertainty in parallax ($\sim$ 3 times larger than the typical parallax uncertainty in our sample), shown as a red triangle in Figure \ref{fig:kinematic cuts}. In total, this procedure provides us a sample comprised of 43 qualifying stars with Hectochelle radial velocities, 36 in L1 and 7 in L2. We also remove 25 stars from the original A22 catalog that did not pass our radial velocity or metallicity cuts. Three relatively faint stars from the A22 catalog with MMT-Hectochelle radial velocities appear fully consistent with the cluster, but were removed from the dynamical sample due to poor spectral quality; they are included in our final catalog but their radial velocities are not used in our dynamical analysis. Our final sample of stars with precisely measured radial velocities is shown in the top panel of Figure \ref{fig:L_plot}, over-plotted on our Theia 456 catalog (referred to as T24 catalog hereafter). Throughout this work, we will separately use both the sample of 43 members with radial velocities as well as all 321 identified members (one of the stars in our sample observed by Hectochelle was not originally in the A22 catalog but nevertheless fit all selection criteria, bringing our Theia 456 catalog to 321 members). An abbreviated catalog of members, including stellar parameters derived via \texttt{MINESweeper} for stars with precise Hectochelle radial velocities, is presented in the appendix in Table \ref{tab:catalog}.

\section{The Two Lobes of Theia 456} \label{sec:2 lobes of Theia 456}

Having selected our sample with follow-up Hectochelle-derived radial velocities, we first analyze the compactness of Theia 456 at earlier times by using astrometric positions, velocities, and Hectochelle radial velocities to integrate stars backwards in time.  We utilize \texttt{gala}, a python package dedicated to integrating galactic orbits \citep{gala}. With the default \texttt{Astropy v4.0} \citep{astropy_v3} parameters, we use the built-in \texttt{MilkyWayPotential}, which combines an NFW halo potential \citep{NFW_1996}, a spherical bulge and nucleus (both represented by Hernquist potentials \citep{Hernquist_1990}), and a multi-component disk model, {which uses a Miyamoto-Nagai potential, for all integrations \citep[for information regarding the disk model, see][]{Bovy_2015}. Note that we consider variations to this model in Section \ref{sec:milky way potential}.

To obtain a qualitative understanding of the evolution of the two lobes, we define a metric $L$ which denotes the 3D spatial distance between the mean $X$, $Y$, and $Z$ positions of L1 and L2 stars. The top panel of Figure \ref{fig:L_plot} shows a schematic of our metric $L$ between the two lobes, and the bottom panel shows $L$ as a function of backwards integration time. The two lobes start roughly 100 pc apart and come closer in space until reaching a local minima 50 Myr ago. They then expand away from each other, reaching a distance of 200 pc at 175 Myr, before coming very close together ($\simeq$ 20 pc) $\simeq$ 250 Myr ago. This latter minima at 250 Myr is consistent with the isochrone and gyrochronology ages reported in A22. 

\begin{figure}
    \begin{center}
    \includegraphics[width=1.0\columnwidth]{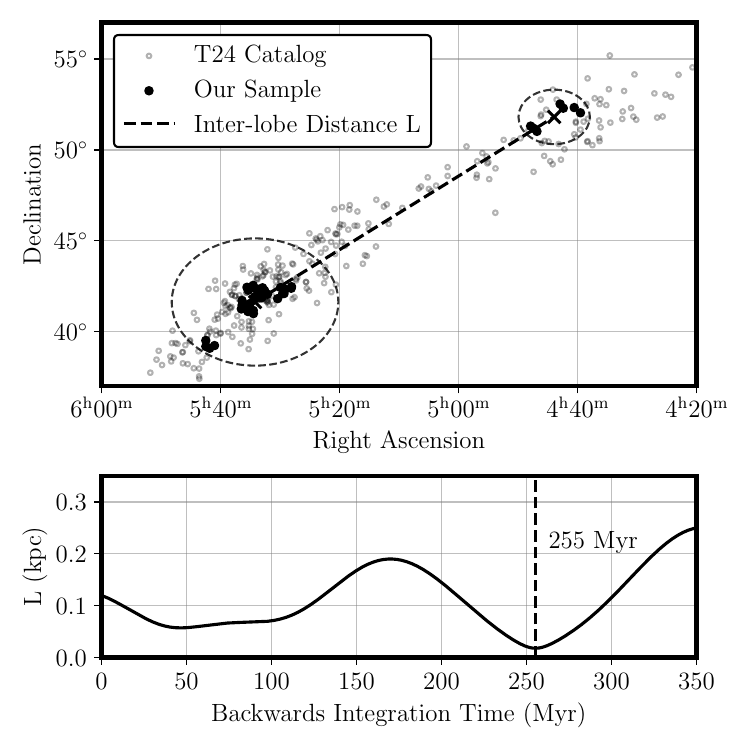}
    \caption{ \textit{Top}: Our sample of stars with Hectochelle-derived radial velocities used for our backwards integrated dynamical analysis (black points), over-plotted on the entire Theia 456 catalog. Stars are segmented into lobes---stars encircled in the southern portion are considered lobe 1 (L1) stars, stars encircled in the northern portion are considered lobe 2 (L2) stars. The dashed line shows the three-dimensional distance $L$ between the two lobes, taken as the distance between the mean X, Y, and Z Galactocentric positions of L1 and L2 stars. \textit{Bottom}: The L1--L2 distance $L$ as a function of backwards integration time within a Milky Way potential. The two lobes nearly overlap in space $\simeq$ 250 Myr ago.}
    \label{fig:L_plot}
    \end{center}
\end{figure}

\begin{figure*}
    \begin{center}
    \includegraphics[width=1.0\textwidth]{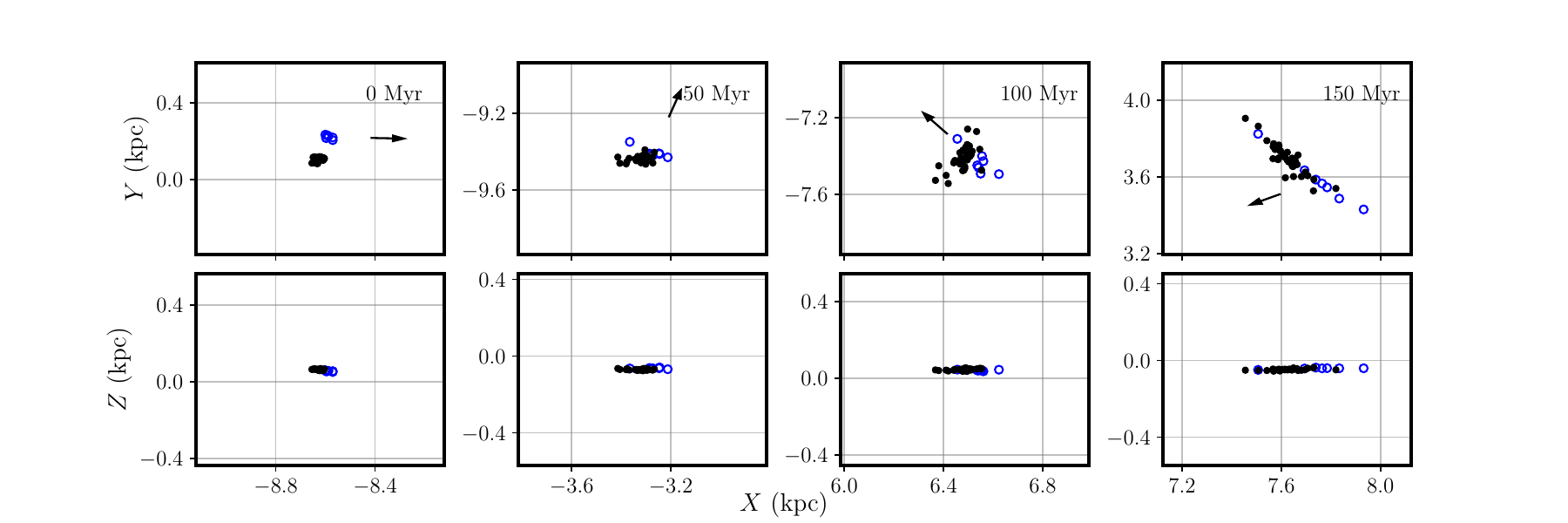}
    \includegraphics[width=1.0\textwidth]{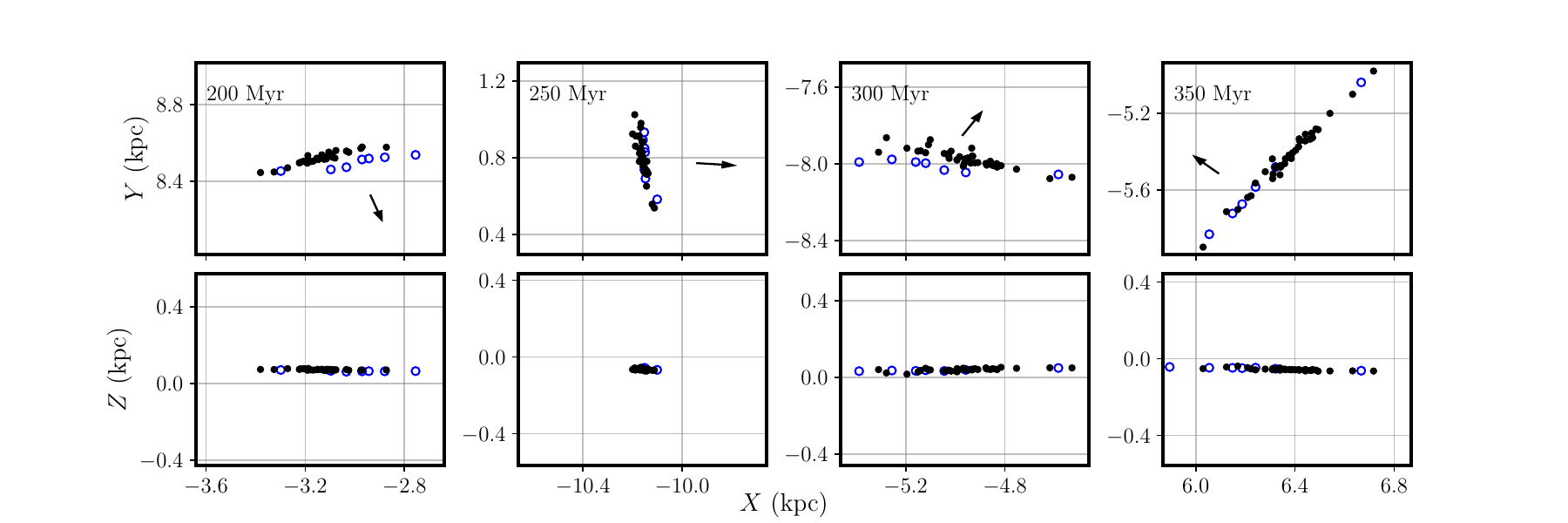}
    \caption{ Slices in the $X$--$Y$ ($1$ kpc $\times$ $1$ kpc) and $X$--$Z$ ($1$ kpc $\times$ $1$ kpc) Galactocentric planes of the positions of Theia 456 stars, backwards integrated at various times under the \texttt{MilkyWayPotential} in \texttt{gala}. Each column is a time-slice of backwards integrated orbits. Arrows in the $X$--$Y$ plane point toward the Galactic center. Closed black circles show L1 stars, open blue circles show L2 stars. The two lobes overlap 250 Myr ago. }
    \label{fig:XY_XZ_slices}
    \end{center}
\end{figure*}

In Figure \ref{fig:XY_XZ_slices}, we show the positions of L1 and L2 stars in the Galactocentric plane at various backwards integrated time-slices. Arrows in the $X$--$Y$ plane point towards the Milky Way center. Here, the overlap between L1 and L2 stars is visible at $50$ and $250$ Myr, and the expansion and contraction of our metric $L$ can be seen as a function of time. In the $250$ Myr panel, we note significant spread along the y-axis. Some fraction of this spread is due to measurement uncertainty in Gaia parallax, and some is due to the natural dynamics of Milky Way orbits; we address the shape drawn out by the error ellipses in parallax and radial velocity in Section \ref{sec:age and scale}. 

Theia 456 is an excellent candidate for the deployment of this method due to the high quality kinematic data at both ends of the structure. By considering the bulk motion of both lobes and not focusing on the individual positions of stars integrated backwards, we are able to reduce the effect of observational uncertainties from the data. Typical astrometric uncertainties for individual Theia 456 stars in $\mu_{\alpha^{\star}}$ ($\simeq$ $0.13\; \rm km~s^{-1}$), $\mu_{\delta}$ ($\simeq$ $0.07\; \rm km~s^{-1}$), $\varpi$ ($\simeq$ $0.023$ mas), and radial velocity ($\simeq$ $0.1\; \rm km~s^{-1}$) translate to a positional uncertainty of roughly 50 pc when integrated backwards in time by 250 Myr, though parallax uncertainties dominate this spread. Theia 456's structure is particularly advantageous; its two distinct lobes, which are quite far apart today ($\simeq$ 120 pc), provide a great deal of leverage when compared with more compact open clusters.

\section{Quantitative Dynamical History} \label{sec: quantitative dynamical history}

\subsection{Statistical Model} \label{sec: statistical model}

To derive direct constraints on the origin of Theia 456, we formulate a Bayesian statistical model that defines the distribution of Theia 456 at birth in an effort to best generate the distribution of 5D Gaia astrometry and Hectochelle radial velocities seen today. We define our model parameters as:
$$ \psi_{0} : \{X, Y, Z, V_X, V_Y, V_Z, \sigma_v, R, T\}$$
where $X, Y, Z$ and $V_X, V_Y, V_Z$ refer to the cluster's mean position and bulk velocity in a Galactocentric coordinate frame at birth, $\sigma_v$ is an isotropic random velocity component given to each model star, $R$ is the initial scale parameter, and $T$ is the cluster age. The model is populated as a 3D Gaussian, such that $R$ refers to the standard deviation in position in all three dimensions, and $\sigma_v$ refers to the standard deviation about the bulk velocity in all three dimensions, making it a one-dimensional velocity dispersion.

\begin{figure*}
    \begin{center}
        \includegraphics[width=1.0\textwidth]{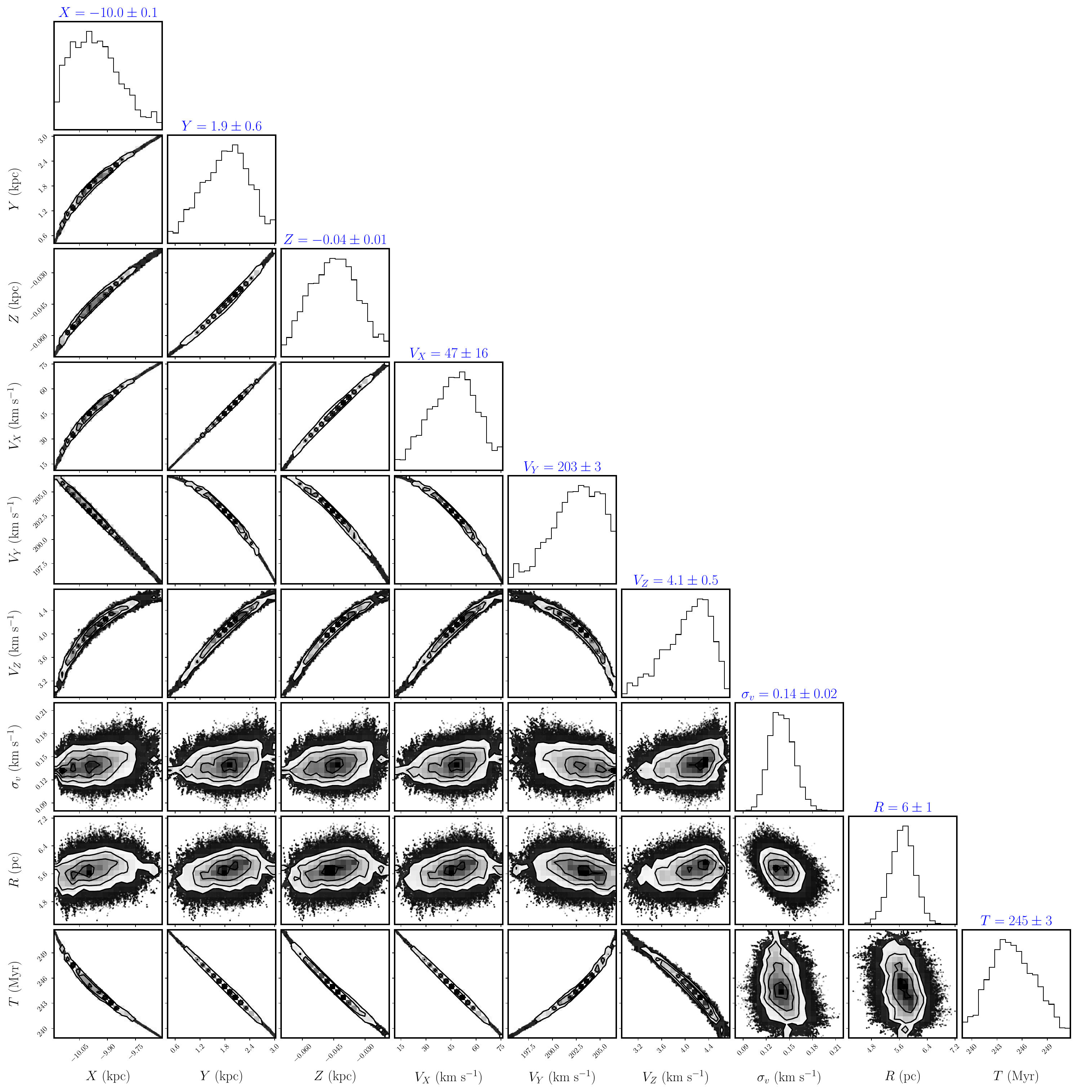}
        \caption{Posterior distributions from our MCMC model describing Theia 456's birth conditions ($\psi_0$), derived via maximization of equation \ref{eq:7}. }
        \label{fig:model parameters}
    \end{center}
\end{figure*}

Given measured observables $x_{f,k}^{\, \prime} = (\alpha, \delta, \varpi, \mu_{\alpha^{\star}}, \mu_{\delta}, V_{\rm rad})$ for the $k$-th star in Theia 456 in its final observed phase space position, we seek to find the following posterior distribution:
\begin{equation}\label{eq:1}
    P(\psi_{0} | \{x_{f}^{\, \prime}\} ) \propto P(\{x_{f}^{\, \prime}\} | \psi_{0} ) P(\psi_{0})
\end{equation}
where $\{ x_{f}^{\, \prime}\}$ refers to the set of all $x_{f,k}^{\, \prime}$. Since stars are observed independently, we can split our conditional likelihood on the r.h.s.:
\begin{equation}
    P(\{x_{f}^{\; \prime}\} | \psi_{0} ) = \prod_{k} P(x_{f,k}^{\, \prime} | \psi_{0} ) 
\end{equation}

Our model ($\psi_{0}$) defines a distribution of initial positions ($x_{i}$), that are then uniquely mapped to final positions ($x_{f}$) via our chosen Milky Way potential, such that $x_{f} = f(x_{i}, \, T)$. We marginalize over the distribution of initial and final positions:
\begin{equation}\label{eq:3}
    P(x_{f,k}^{\, \prime} | \psi_{0} ) = \int P(x_{i}, x_{f}, x_{f,k}^{\, \prime} | \psi_{0} ) \, d x_{i} \, d x_{f} \equiv \int I \, d x_{i} \, d x_{f} 
\end{equation}
where the integrand simplifies as:
\begin{eqnarray}
        I &=& P(x_{f,k}^{\, \prime} | x_{i}, x_{f}, \psi_{0} ) P(x_{f} | x_{i}, \psi_{0}) P(x_{i} | \psi_{0}) \nonumber \\ 
         &=& P (x_{f,k}^{\, \prime} | x_{f}) P(x_{f} | x_{i}, T)  P(x_{i}| \psi_{0})
\end{eqnarray}

where we utilize the fact that $x_{f,k}^{\, \prime}$ depends only on $x_{f}$, and $x_{f}$ depends only on $x_{i}$ and $T$. Since we uniquely map the distribution of initial positions to final positions, we have $ P(x_{f} | x_{i}, T) = \delta(x_{f} - f(x_{i}, T))$. This simplifies equation \ref{eq:3} as:
\begin{equation}\label{eq:5}
    \begin{array}{l}
        P(x_{f,k}^{\, \prime} | \psi_{0} ) = \int P(x_{f,k}^{\, \prime} | x_{f}^{\star}) P(x_{i}| \psi_{0}) \, d x_{i} \\
        P(x_{f,k}^{\, \prime} | \psi_{0} ) = \int \mathcal{N}(x_{f}^{\star}; x_{f,k}^{\, \prime}, \sigma^{ \prime}) P(x_{i}| \psi_{0}) \, d x_{i}
    \end{array}
\end{equation}
where $x_{f}^{\star} = f(x_{i}, T)$ in ICRS coordinates, and $\sigma^{\prime}$ refers to the observational uncertainties on $x_{f,k}^{\, \prime}$, which we keep consistent across all $K=321$ stars. We discuss the reasoning for this later in this section. Note that we include radial velocities for the 43 stars in our sample outlined in Section \ref{sec:dynamical members}, and only use Gaia astrometry for the remaining 278 stars. 

We approximate the integral in equation \ref{eq:5} as a sum such that:
\begin{equation}
    P(x_{f,k}^{\, \prime} | \psi_{0} ) \approx \frac{1}{N} \sum_{n} \mathcal{N}(x_{f,n}^{\star}; x_{f,k}^{\, \prime}, \sigma^{ \prime})
\end{equation}
where $x_{f,n}^{\star} = f(x_{i,n}, T)$ and $x_{i,n}$ is drawn from the distribution $P({x_{i} | \psi_{0}})$.

Equation \ref{eq:1} simplifies, in $\log$-space to:
\begin{multline}\label{eq:7}
    \log P(\psi_{0} | \{x_{f}^{\, \prime} \} ) \propto \sum_{k} \log \bigg[ \frac{1}{N} \sum_{n} \mathcal{N}(x_{f,k}^{\, \prime}; x_{f,n}^{\star}, \sigma^{ \prime}) \bigg] \\ + \log P(\psi_{0})
\end{multline}
where we apply flat priors on all model parameters, though our Bayesian approach allows for easy adjustments if one wants to incorporate priors, such as an isochrone age, or additional model parameters.

\begin{figure*}
    \begin{center}
    \includegraphics[width=1.0\textwidth]{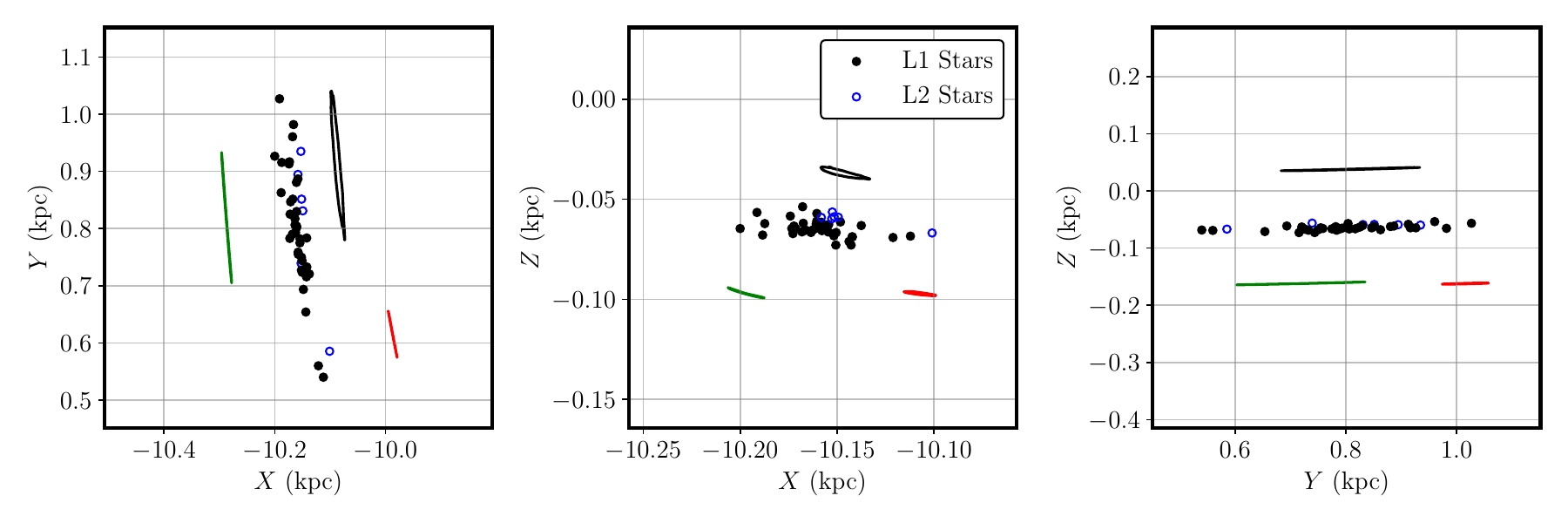}
    \caption{Theia 456, backwards integrated by $T = 250$ Myr. Black contours show the size and shape of the error ellipses of one star with typical uncertainties, with full propagation of 5D astrometric and radial velocity uncertainties. Green contours show error ellipses for the same star, 
    but only considering uncertainties in parallax. Red contours only consider uncertainties in proper motion (both $\mu_{\alpha^{\star}}$ and $\mu_{\delta}$). Typical radial velocity uncertainties produce negligibly small error ellipses. Contours are drawn at the $2 \sigma$ level containing 95\% of the data.}
    \label{fig:integrated error ellipses}
    \end{center}
\end{figure*}

To find optimum model parameters, we employ $N = 10,000$ model stars, and use \texttt{emcee} \citep{emcee_ref_2013} to maximize our posterior via a Markov Chain Monte Carlo (MCMC) simulation with 64 walkers for 75,000 steps. To minimize effects of stochasticity, we take the following steps. First, we increase the uncertainties in our 5D Gaia astrometric data and Hectochelle radial velocities, and apply these uncertainties consistently to all $K=321$  stars in our sample. We choose uncertainties of: 
$$
\sigma^{\prime} = (\sigma_{\alpha}, \sigma_{\delta}, \sigma_{\varpi}, \sigma_{\mu_{\alpha^{\star}}}, \sigma_{\mu_{\delta}}, \sigma_{V_{\rm rad}}) 
$$
$$
\sigma^{\prime} = (1^{\circ}, \, 1^{\circ}, \, 0.1 \; \rm{mas}, \, 0.15 \; \rm{mas \; yr^{-1}}, \, 0.15 \; \rm{mas \; yr^{-1}}, \, 1 \; \rm{km~s^{-1}})
$$
so that the observable data points have overlapping error bars, but do not significantly overshoot the width of the distribution in phase space. We also ignore covariances in equation \ref{eq:7} because our newly defined uncertainties dominate. This is a reasonable step to take because we are not interested in precisely reproducing individual stars' observables; rather we are interested in finding model parameters that reproduce the size, shape, and location of the distribution of stars in Theia 456. Attempting to fit model stars to the extremely precise Gaia and Hectochelle uncertainties leads to numerical convergence only with an unreasonably large value of $N$. To maintain detailed balance, we draw one set of $\{ x_{i,n} \}$ at the beginning of our simulation, and shift and re-scale this distribution according to the new model parameters at each step. This ensures that returning to the same model parameters at a later step reproduces the same likelihood. To confirm that our chosen value of $N$ is sufficiently large, we perform convergence tests and find that our results are robust to different initial random draws with our choice of $\sigma^{\prime}$ and $N$.

\subsection{Velocity Dispersion, Birth Size, and Age of Theia~456} \label{sec:age and scale}

Figure \ref{fig:model parameters} shows our resulting model parameters and their covariances. Of particular importance to us, we find: 
$$\sigma_{v_{\rm opt}} = 0.14 \pm 0.02 \; \;  \rm km~s^{-1} $$   
$$R_{\rm opt} = 6 \pm 1 \; \; \rm pc $$ 
$$T_{\rm opt} = 245 \pm 3 \; \; \rm Myr$$ 
where all quoted uncertainties are statistical. Since we generate positions by populating a Gaussian in 3D, the half-mass radius, assuming that mass is randomly distributed at birth, is $R_{h, \rm opt} \approx 1.5 \, R_{\rm opt} \approx 9$~pc. We note that our model does not take any consideration of natal gas, and that derived model parameters likely represent Theia 456 in its post-embedded phase, potentially making the true age of the structure marginally older \citep[for an analysis on the timescale at which natal gas is expelled, see][]{Dinnbier_2020_Gas}. Our derived age is consistent with both the isochrone and rotation period age derived in A22, as well as the ages of individual stars derived via spectral fitting reported in Table \ref{tab:catalog}.

In Figure \ref{fig:integrated error ellipses}, we show backwards integrated positions of Theia 456 at $T = 250$ Myr, this is essentially a zoom-in of the 250 Myr panels in Figure \ref{fig:XY_XZ_slices}. We note significant spread ($\sim$ 500 pc) along the y-axis. Contours represent the size and shapes of 2$\sigma$ error ellipses of a single star with typical astrometric and radial velocity uncertainties for our sample. The black contours, which span roughly half of the entire spread along the y-axis, take into account full error propagation of the 5 astrometric properties and radial velocity. Green and red contours show error ellipses due only to parallax and proper motion, respectively. Typical uncertainties ($\sim 0.1 ~\rm km~s^{-1}$) in Hectochelle-derived radial velocity sweep out negligibly small error ellipses. The error ellipses swept out by parallax uncertainties cover almost the entire length of the full error ellipse, indicating that these uncertainties are the predominant cause of the observed spread about the y-axis. 

Figure \ref{fig:integrated error ellipses} further showcases the difficulties of backwards modeling a cluster of this age, as the propagation of errors leads to an unavoidable spread in birth positions. With currently available data, deriving precise constraints about the birth position and velocity of any individual star is fruitless. Nevertheless, the close match between the error ellipses and Theia 456's elongated distribution, suggests the derived birth positions are an artifact of the propagation of observational uncertainties under our Milky Way model; Theia 456's shape at birth is consistent with that of a sphere as described by our generative forward model.

\subsection{Posterior Predictive Checking} \label{sec: posterior predictive checking}

\begin{figure*}
    \begin{center}
        \includegraphics[width=1.0\textwidth]{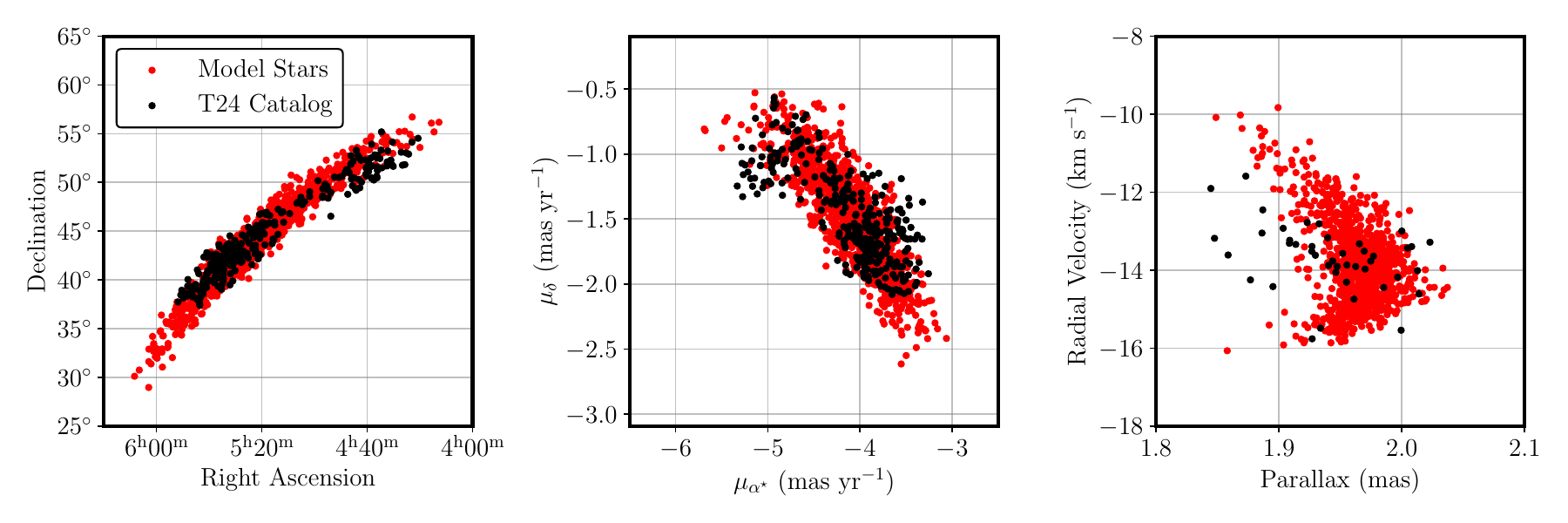}
        \caption{Theia 456 observables (black points), compared with forward integrated model stars generated using the optimal model parameters shown in Figure \ref{fig:model parameters} (red points). In position and proper motion space, we show all 321 stars in Theia 456, and show only stars with Hectochelle-derived radial velocities in parallax-radial velocity space.}
        \label{fig:model stars phase space}
    \end{center}
\end{figure*}

While our model clearly favors an age of $T_{\rm opt} = 245$ Myr, size of $R_{\rm opt} = 6$ pc ($R_{\rm h, opt} = 9$ pc), and velocity dispersion of $\sigma_{v_{\rm opt}} = 0.14\; \rm km~s^{-1}$, we wish to check if these parameters provide a reasonable representation of the observables when forward integrated. To qualitatively assess the validity of these parameters, we generate 10,000 model stars from the derived birth parameters, and integrate them forward for a time $T_{\rm opt}$. These forward integrated model stars are compared with observables in 6D phase space in Figure \ref{fig:model stars phase space}. We see good agreement in all phases of 6D parameter space, and are able to reproduce the bulk observables of Theia 456. Particularly, the observed scale and direction of Theia 456's morphology in position and proper motion space corresponds with the predicted dispersion of a loosely bound open cluster via  Galactic tides. Our derived birth conditions for Theia 456 indicate that the cluster was never strongly bound, at least in its post-embedded phase, implying that internal cluster dynamics may not be the dominant dynamical effect in such low density expanding structures. However, our spherical model does not perfectly reproduce what we see on the sky today. Theia 456 appears on the sky as two overdense lobes connected by a diffuse bridge, whereas our generative model produces an elongated structure with a monomodal distribution. This indicates that a spherically symmetric ball is not a perfect physical description of Theia 456 at birth, and that initial substructure may play a role; it may have been a more elongated structure, or a more dumbbell-like structure to produce the disperse bridge seen. As an example, \cite{Briceno-Morales_2023} show that Upper Scorpius contains significant kinematic substructure, and \cite{Arunima_2023} backtrack orbits of stars in Upper Scorpius to determine dynamical ages for the individual substructures, showing that unbound stars can preserve initial substructure. We leave a more dedicated study, with a focus on initial substructure, to reproduce these particular features of Theia 456 for future work. We also note that we are able to accurately reproduce the spread in all phase space dimensions but parallax, which may indicate that uncertainties on Gaia parallaxes are somewhat underestimated for our sample. 

To check the validity of $\sigma_{v_{\rm opt}}$, $R_{\rm opt}$, and $T_{\rm opt}$ individually we perform the following procedure. First, we integrate our sample backwards by a time $T$ (which may be different than $T_{\rm opt}$) which forms the initial position for our trial solution. Next, we generate a cluster of 10,000 model stars with $\sigma_{\rm pos} = R$ at the mean birth position of our backwards integrated sample and initialize their velocities with the cluster bulk velocity and isotropic random component $\sigma_v$. Finally, we integrate the stars drawn from this trial solution forward in time for three different velocity dispersions, ages, and birth sizes.

\begin{figure*}
    \begin{center}
    \includegraphics[width=1.0\textwidth]{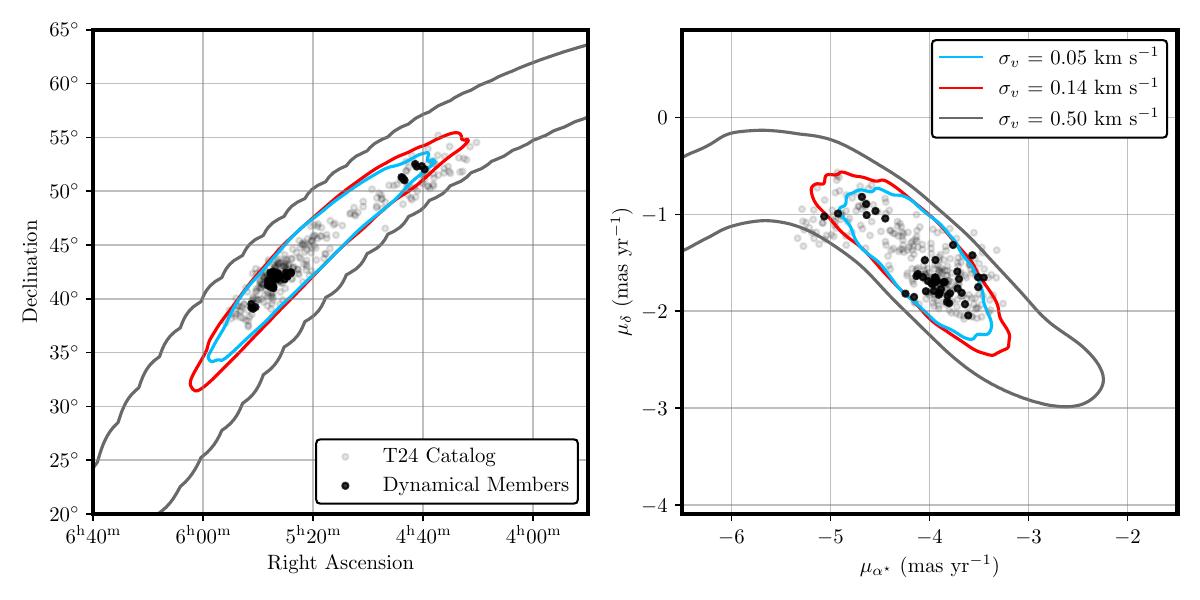}
    \caption{Comparison of Theia 456's phase space at the present epoch with that of a sphere forward-integrated under the \texttt{MilkyWayPotential}, centered at the mean position given by backwards integrating dynamical members to time $T_{\rm opt}$, with a radius of $R_{\rm opt}$. Each contour contains 95\% of model stars forward integrated, drawn from a Maxwell--Boltzmann distribution at different values of $\sigma_v$. Black and grey points denote our sample of stars with Hectochelle radial velocities and the entire Theia 456 catalog, respectively. Our model (the red contour) prefers a relatively low velocity dispersion at birth; $\sigma_{v_{\rm opt}} = 0.14\; \rm km~s^{-1}$, larger velocity dispersions produce overly diffuse position and proper motion spaces.}
    \label{fig:forward_model_MB_vels}
    \end{center}
\end{figure*}

In Figure \ref{fig:forward_model_MB_vels}, we compare the observed sky positions and proper motions of our sample today with those produced by our generative model at different values of $\sigma_v$, but still using our fiducial model parameters $R_{\rm opt}$ and $T_{\rm opt}$. Contours contain 95\% of the simulated data. Even a modestly larger velocity dispersion leads to an overly disperse stream, indicating that Theia 456's initial velocity dispersion is tightly constrained to be no more than a few $\sim 0.1 \; \rm km~s^{-1}$. We note that this velocity dispersion is lower than what is typically found in young star-forming regions \citep[for example, NGC 1333 was found to have $\sigma_v = 0.92 \rm ~km~s^{-1}$;][]{Foster_2015}.

\begin{figure*}
    \begin{center}
    \includegraphics[width=1.0\textwidth]{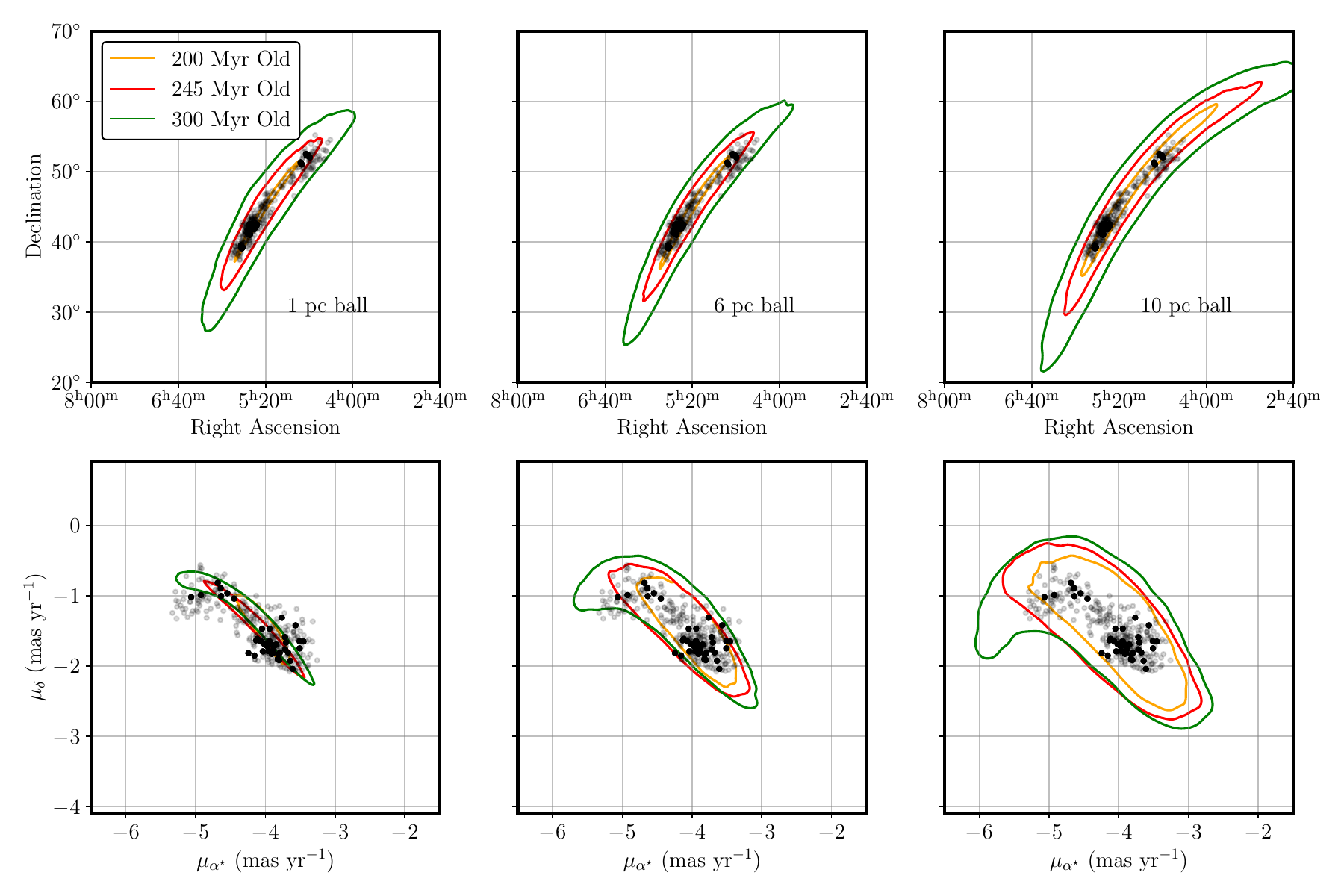}
    \caption{Mock phase space of Theia 456 at the present epoch from various initial conditions, compared with the observed phase space of our sample with precise radial velocities (black points) and the entire Theia 456 catalog (gray points). Contours contain 95\% of the data and are drawn at different birth ages, while columns refer to different birth sizes. Red contours in the middle column most closely represent our best fit model.}
    \label{fig:forward_model_MB_ages_sizes}
    \end{center}
\end{figure*}

Finally, Figure \ref{fig:forward_model_MB_ages_sizes} shows the positions and proper motions predicted by our forward models at different initial birth times and scales, at our derived velocity dispersion $\sigma_{v_{\rm opt}}$, compared with the observed phase space of Theia 456 stars today. Each column represents a different initialization size, while each contour represents a different age. A $10$ pc ball clearly produces a too-extended stellar distribution in proper motion space, independent of birth time. For the older age of $300$ Myr, the same is true, as the position distribution is too extended compared with our observed sample. In general, sizes and ages larger than our optimal model parameters are overly influenced by Galactic tides and appear more disperse in phase space than what is observed in Theia 456 today. Decreasing the age and size leads to the opposite effect; Galactic tides are not strong enough to reproduce the size of the distribution observed in phase space today. We conclude that indeed, the distribution of stars in position and proper motion space is a sensitive tracer of the cluster's birth conditions, as our relatively simple model accurately reproduces the main features of Theia 456 as observed today.

\section{Discussion}\label{sec:discussion}

\subsection{Sensitivity to the Milky Way Potential}\label{sec:milky way potential}

Up until this point, we have only considered the effect of observational uncertainties and how they propagate backwards in time as a function of the Milky Way potential. To further test the robustness of our model predictions, we also consider variations in the potential itself based on our imperfect knowledge of its functional form. Specifically, we modify the characteristic mass $M$ in the Miyamoto-Nagai \citep{Miyamoto_1975} disk-component of the \texttt{MilkyWayPotential}, shown below: 
\begin{equation}\label{eq:miyamoto-nagai}
\Phi(R, z) = - \frac{GM}{\sqrt{R^2 + (a + \sqrt{z^2 + b^2})^2}}
\end{equation}
and compute the differences in model outputs in $\sigma_v$, $R$, and $T$ under these modifications. The default value is $M_{\rm def} = 6.8 \times 10^{10}$ $\rm M_{\odot}$. We consider only variations in the disk component of the model, and not the bulge, nucleus, or halo, as we find that the disk component of the potential is the dominant dynamical driver for a thin-disk structure like Theia 456. We consider variations in $M$ such that the computed circular velocity of the Sun, at an assumed Galactocentric radius of $8.12$ kpc, is reasonably varied. We do not consider variations in $a$ and $b$, as variations in $a$ produce very similar effects as variations in $M$, and even large variations in $b$ produce negligible differences in model parameters. At the fiducial disk parameters in \texttt{gala}, $V_{\rm circ} = 231.52\; \rm km~s^{-1}$; we modify $M$ in equation \ref{eq:miyamoto-nagai} by 20\% in either direction, such that $V_{\rm circ}$ ranges from $[219, 244]$ $\rm km~s^{-1}$, a reasonable encapsulation of the current constraints on $V_{\rm circ}$ \citep{Eilers_2019}. Table \ref{tab:milky way uncertainties} shows the outputted model parameters due to these modifications. Our model constraints on cluster age are quite robust; reasonable variations to the disk component of the potential produce only a modest $5\%$ maximum change in $T$. The velocity dispersion and radius are less robust to variations in the Galactic potential; $\sigma_v$ and $R$ vary by up to $64\%$ and $50\%$, respectively. While there is some evidence that the Milky Way disk potential has some azimuthal variations and is potentially time-dependent \citep{Antoja_2018, Amores_2017}, Theia 456 lies in the thin disk in the Solar Neighborhood, where the potential is particularly well constrained. Even if our chosen \texttt{MilkyWayPotential} is an imperfect representation of our Galaxy, we expect that the models provided in Table \ref{tab:milky way uncertainties} span a reasonable range of outcomes.

\begin{deluxetable}{ccccc}
 \tablecaption{Model Outputs for Variations in \texttt{MilkyWayPotential} \label{tab:milky way uncertainties}}
 
 \tablehead{
    \colhead{$\frac{\displaystyle M}{\displaystyle M_{\rm def}}$ } & 
    \colhead{$V_{\rm circ}$ (km s$^{-1}$)} &
    \colhead{$\sigma_{v}$ (km s$^{-1}$)} & 
    \colhead{$R$ (pc)} &
    \colhead{$T$ (Myr)}
    }
\startdata \vspace{0.1cm}
0.8 & 219  & $0.23 \pm 0.01$ & $3 \pm 1$  & $257 \pm 2$\\ \vspace{0.1cm}
\textbf{1} & $\mathbf{232}$  & $\mathbf{0.14 \pm 0.02}$ & $\mathbf{6 \pm 1}$ & $\mathbf{245 \pm 3}$ \\ \vspace{0.1cm}
1.2 & 244  & $0.11 \pm 0.01$ & $5 \pm 1$  & $252 \pm 2$\\
\enddata
\tablenotetext{}{Quoted statistical uncertainties are 1$\sigma$. $V_{\rm circ}$ is the circular velocity of the Sun at that given disk mass.}

\end{deluxetable}

While we assume an axisymmetric and time-independent potential; more complex considerations, such as perturbations from the spiral arms and giant molecular clouds (GMCs), require a more focused work on the modeling of the Milky Way potential. Such baryonic substructure in the disk has been shown to disrupt passing star clusters \citep{Gieles_2006, Gieles_2007, Gieles_2016}. These perturbative features, while not necessary to reproduce the bulk distribution seen today, may be the origin of the particular features of Theia 456, including its diffuse bridge. The bridge is somewhat reminiscent of observed gaps in the GD-1 and Palomar 5 halo streams which have been previously explained through the gravitational interaction of stars with Milky Way substructure \citep{Bonaca_2019, Pearson_2017}.  We leave the issue of whether any such encounter with substructure included in a more realistic Milky Way potential has a significant effect on the derived dynamical age, radius, and velocity dispersion for future work.

\subsection{The Stellar Population of Theia 456} \label{sec:initial mass estimate}

In order to obtain an order of magnitude estimate for the total number of stars and stellar mass of Theia 456, we extrapolate via a Kroupa initial mass function \citep{Kroupa_2001}. To solve for the normalization constant, we use the fact that Gaia is essentially complete for $G < 17$ \citep{Gaia_DR3_documentation}. We solve for the normalization constant $N_0$ using:
\begin{equation}
n_{\rm obs} = N_0 \int_{M_{\rm min}}^{\infty} m^{\alpha} dm
\end{equation}
where $n_{\rm obs}$ is the number of Theia 456 stars observed with $G < 17$, $M_{\rm min}$ is the minimum mass in this fully covered region, and $\alpha$ is the Kroupa IMF power law index ($\alpha = -0.3$ for $ 0.01 \; \text{M}_{\odot} < m < 0.08$ M$_{\odot}$, $\alpha = -1.3$ for $ 0.08 \; \text{M}_{\odot} < m < 0.5$ M$_{\odot}$, $\alpha = -2.3$ for $ m > 0.5$ M$_{\odot}$). 

We utilize the FLAME (Final Luminosity and Age Estimator) module of the astrophysical parameters inference system (Apsis) chain \citep{Bailer-Jones_2013, Fouesneau_2023} to determine $M_{\rm min} = 0.57$ M$_{\odot}$, simply taking the smallest mass from the FLAME module for Theia 456 stars with $G < 17$. Having solved for our normalization constant $N_0$, we estimate the total number and mass of Theia 456 by invoking continuity of the IMF at crossover points to solve for normalization constants in those regimes. Under these assumptions, we find $n_{\rm total} \simeq 2100$ and $m_{\rm total} \simeq 900 \; \rm M_{\odot}$. Note that this mass is roughly twice the mass reported by \cite{Bai_2022}, as they only consider stars brighter than $G = 18$. At our derived mass and half-mass radius, our results suggest that Theia 456 was a low density ($\approx 0.2 ~ \rm M_{\odot}~pc^{-3}$) open cluster at birth. As a comparison, the Pleiades has a current day mass ($\approx 740$ M$_{\odot}$) and half-mass radius ($\approx 3.66$ pc), leading to a characteristic density of $\approx 1.8$ $\rm M_{\odot}~pc^{-3}$ \citep{Moraux_2004}. Given the possibility that the Pleiades is expanding \citep{Converse_2010}, its density at birth may have been even higher. Although we account for the magnitude limitations of Gaia, incompleteness due to limitations of the \texttt{HDBSCAN} algorithm used to identify members of Theia 456, particularly near the edges of the structure, may drive this density to be slightly higher; we discuss these limitations further in Section \ref{sec:Future Dispersion of Theia 456}.

\subsection{Self-Gravity} \label{sec: self gravity}

In our derivation of model parameters to describe Theia 456 at birth, we have so far ignored any potential effects of self-gravity between stars. The predominant driver of the early dynamical evolution of stars after their gas embedded phase comes from the potential of the Milky Way and the cluster's own self-gravitational interactions. Even when gas is present, it does not play a major role in affecting the dynamical state of the young cluster \citep{Sills_2018}, though its expulsion \citep[i.e., the infant mortality paradigm;][]{Lada_2003, Pellerin_2007} is dynamically important, especially in the formation of tidal tails \citep{Dinnbier_2020}. However, in loosely bound open clusters and stellar streams, self-gravity may not play a major role in the bulk dynamical evolution of the system, due to large inter-stellar separations. 

After its initial embedded phase, the acceleration that a star feels is equivalent to:
\begin{equation}
    \vec{a}_{\rm star} = -\nabla \Phi_{\rm Milky Way} - \nabla \Phi_{\rm Cluster}
\end{equation}
As a simple order-of-magnitude estimate, we take $\Phi_{\rm Cluster}$ to be that of a Plummer sphere \citep{Plummer_1911}, i.e. $\Phi_{\rm Cluster}(r) = -G M / \sqrt{r^2 + a^2}$ where $M$ is the total mass of the cluster, $r$ is the distance to the cluster center, and $a$ is the characteristic scale radius. The maxima in the potential gradient $\nabla \Phi_{\rm Cluster}(r)$ occurs at $r = a / \sqrt{2}$, such that $\nabla \Phi_{\rm Cluster}(a/\sqrt{2}) \approx GM / a^2$. For our cluster mass ($M \approx 900$ M$_{\odot}$) and scale ($a \approx 7$ pc, derived using the fact that $R_{h, \rm opt} \approx 1.3 a$), the resulting characteristic acceleration due to the cluster's self gravity is $\sim 10^{-10}$ cm s$^{-2}$. The gradient in our model Milky Way potential at the predicted birth position of Theia 456 is $\sim 10^{-8}$ cm s$^{-2}$. 

To further justify our exclusion of self-gravity, we initialize a model run with our fiducial parameters found in Section \ref{sec:age and scale} and run it until present day without the inclusion of self-gravity. We populate the run with $n_{\rm total} \simeq 2100$ stars, and assign each star a mass drawn from a Kroupa IMF\footnote{\url{https://github.com/keflavich/imf}} such that $m_{\rm total} \simeq 900$ M$_{\odot}$. At each timestep, we separately compute the acceleration each star feels due to the Milky Way potential, and due to the sum of gravitational forces from other stars.\footnote{\url{https://github.com/mikegrudic/pytreegrav}} In Figure \ref{fig:stellar_accelerations}, we compare the two accelerations, (taking the median value as a representation of the gravitational interactions between stars). The typical acceleration due to the Milky Way potential is orders of magnitude higher than the acceleration due to self-gravity throughout the forward run, indicating that self-gravity can safely be ignored when studying the global dynamical properties of Theia 456. 

\begin{figure}
    \begin{center}
    \includegraphics[width=1.0\columnwidth]{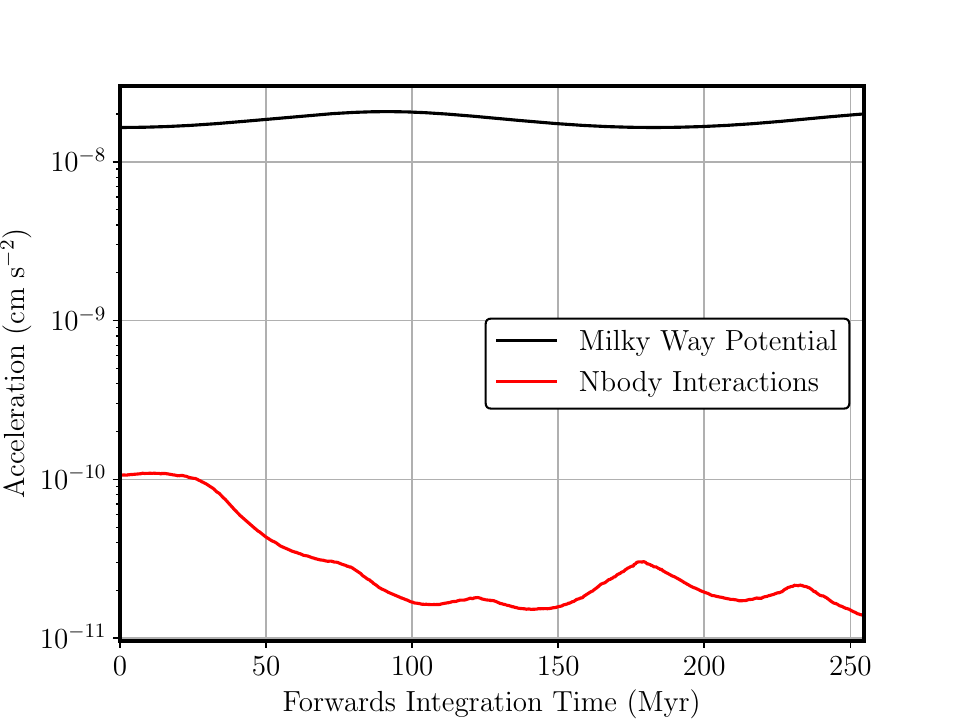}
    \caption{Comparison of the typical acceleration that a model star in Theia 456 feels solely due to the Milky Way potential (black line) and due to the sum of Nbody interactions (red lines). Shown is the median (for 2166 stars) magnitude of the acceleration at each timestep. The Milky Way potential is the dominant driver for the acceleration of stars in Theia 456. }
    \label{fig:stellar_accelerations}
    \end{center}
\end{figure}

While we can ignore self-gravitational interactions in our orbital integrations, we again do not rule out that our optimal model parameters may describe Theia 456 after dynamical relaxation and not strictly at birth; at these earlier times, dynamical interactions, especially those between massive binaries, could indeed be important for the cluster's evolution \citep{Torniamenti_2021}. We also note that Figure \ref{fig:stellar_accelerations} shows only the typical accelerations due to self-gravitational interactions. Close encounters between individual stars may be dynamically important to accurately reproduce the position and velocity of individual stars, which is outside of the scope of this work. We also do not rule out the possibility that there is a small bound core in the densest portion of Theia 456. \cite{Hunt_2024} classify COIN-Gaia-13 as a bound open cluster by determining that the cluster has a valid Jacobi radius. However, only $\sim  20\%$ of stars in their catalog lie within that Jacobi radius, meaning the bulk of Theia 456 members are unbound, which supports our findings that the observed phase space morphology is a product of the Galactic tides, and not internal cluster dynamics.

\subsection{Tracing Birth Conditions using Stellar Positions Today}\label{sec:tracing birth properties}

The idea that stars typically form in bound, dense clusters has  been somewhat challenged as of late \citep[e.g.,][]{Gouliermis_2018, Ward_2020}, favoring the hierarchical collapse of gas across a wide range of surface densities. This picture implies that the invocation of gas expulsion and/or dynamical relaxation may not be necessary to unbind clusters and explain the wealth of observed, unbound gas-rid associations within the Milky Way \citep{Wright_2022}. Under this paradigm, for sufficiently low density structures, the Milky Way tides would play a critical role in the dissolution of the cluster. This picture of star formation occurring at low density scales may apply to Theia 456, explaining its relatively low stellar density ($\approx 0.2 ~ \rm M_{\odot}~pc^{-3}$), though quantifying this is outside of the scope of this work. If one can constrain the size and initial velocity distribution of a stellar stream, metallicity gradients in position today may be indicative of metallicity gradients in the parent molecular cloud. Stellar streams could then provide a probe of the giant molecular clouds from which stars are born. This motivates high-resolution spectroscopic observations to obtain precise chemical abundances of stars to further explore this correlation.

Stellar streams may also pose a powerful tool to study mass segregation; whether mass segregation is primordial or a dynamical effect is still unclear \citep{Parker_2015, Dib_2018}. If one can constrain the initial size and velocity distribution of a stellar population, then the presence (or lack thereof) of mass segregation in streams could help answer this question. In the case of Theia 456, our model predicts $\sigma_{v_{\rm opt}} = 0.14\; \rm km~s^{-1}$, while the virial theorem for a cluster of $m_{\rm total} \approx 900$ M$_{\odot}$ and $R_{h} \approx 9$ pc predicts a velocity dispersion of $\sigma_{v} \approx 0.2\; \rm km~s^{-1}$, using $\sigma_v = \sqrt{GM/\eta R_h}$ with $\eta \approx 10$ as an approximation for a Plummer sphere \citep[see][]{Portegies_2010}. We caution against over-interpreting the similarity in these velocities, as Theia 456 likely had very different conditions at birth, in its embedded state. It is possible, however, that such low-density structures may begin to dissolve on a shorter timescale than their crossing time in their embedded state, indicating that any observed mass segregation is primordial. Precise and accurate mass determinations for a number of stellar streams, coupled with simulations of young low-density clusters undergoing gas expulsion \citep[e.g.][]{Pfalzner_2013}, would be necessary to test this hypothesis.

In Figure \ref{fig:RA_DEC_birth}, we explore the potential correlation between birth position and velocity and the observed phase space today. Using our derived age and birth scale, we forward-integrate 1,000 model stars in the manner described above for three different velocity dispersions ($\sigma_v = 0.14, \; 0.25, \; 0.50\; \rm km~s^{-1}$), and color-code points by their cylindrical Galactocentric birth position (top 3 rows) and birth velocities relative to the bulk velocity of the cluster (bottom 3 rows). For sufficiently low thermal velocities, we see a clear correlation between initial Galactocentric radius and the position on the sky today, visible in the top left panel. Somewhere between $\sigma_v = 0.25\; \rm km~s^{-1}$ and $\sigma_v = 0.50\; \rm km~s^{-1}$, this gradient in initial position begins to obscure; and a correlation between the initial angular velocity and position on the sky today emerges, visible in the bottom middle panel. Qualitatively, this correlation is expected. For low thermal velocities, the tidal disruption timescale, which we approximate as $\Delta t_{\rm tidal} = \sqrt{2 R_{h, \rm opt} / a_{\rm tidal} }  \approx 2 $ Myr, is significantly shorter than the crossing timescale, which at our derived half-mass radius and velocity dispersion is $\approx 50$ Myr. The opposite occurs for higher thermal velocities; stars with large relative angular velocities move quickly to one side of the cluster before tidal forces can disrupt it. At our derived birth parameters (top row), we see agreement with the claim made by \cite{Bai_2022}; it does appear that Theia 456 is dissolving into the field via differential rotation caused by the Galactic tides.

\begin{figure*}
    \begin{center}
    \includegraphics[width=1.0\textwidth]{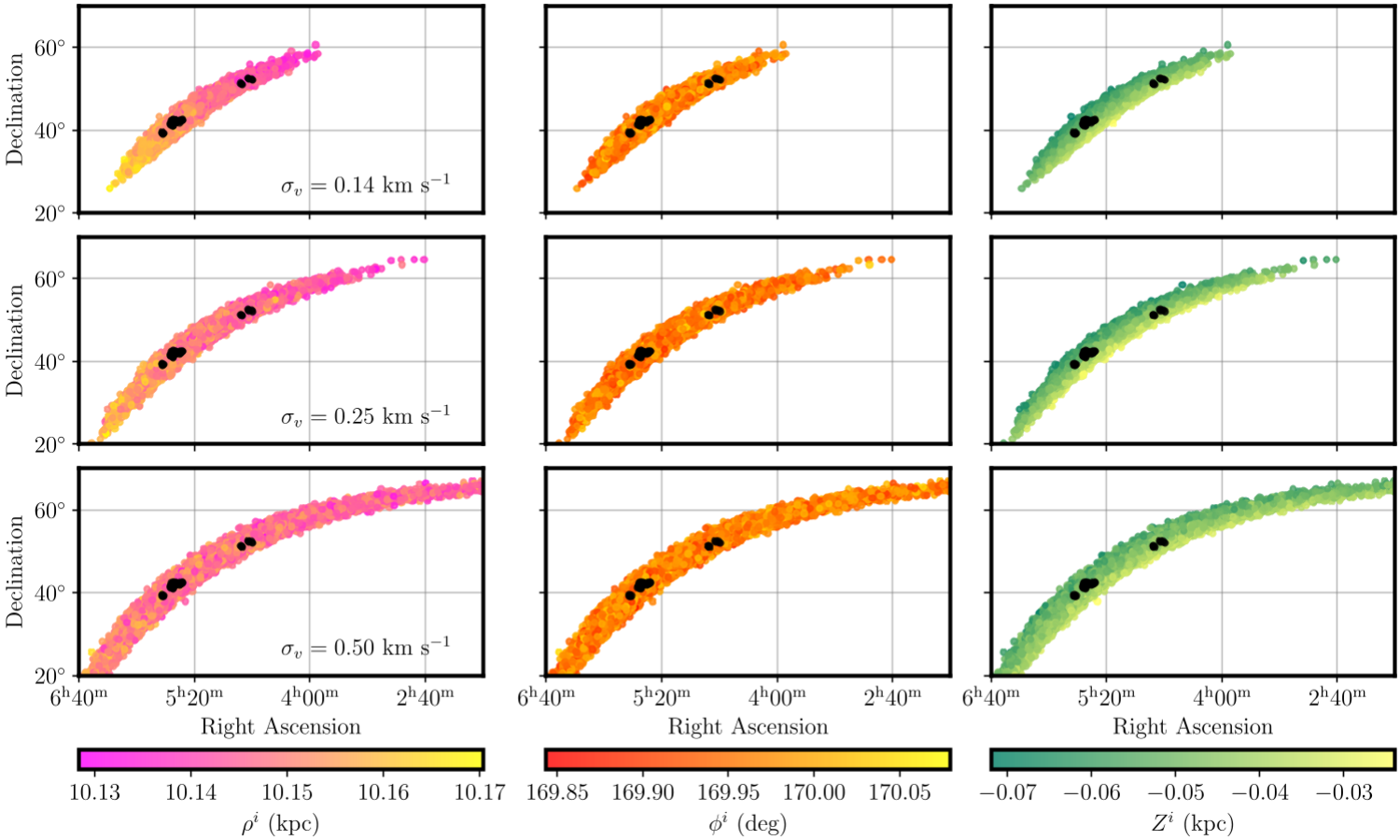}\vspace{0.5cm}
    
    \includegraphics[width=1.0\textwidth]{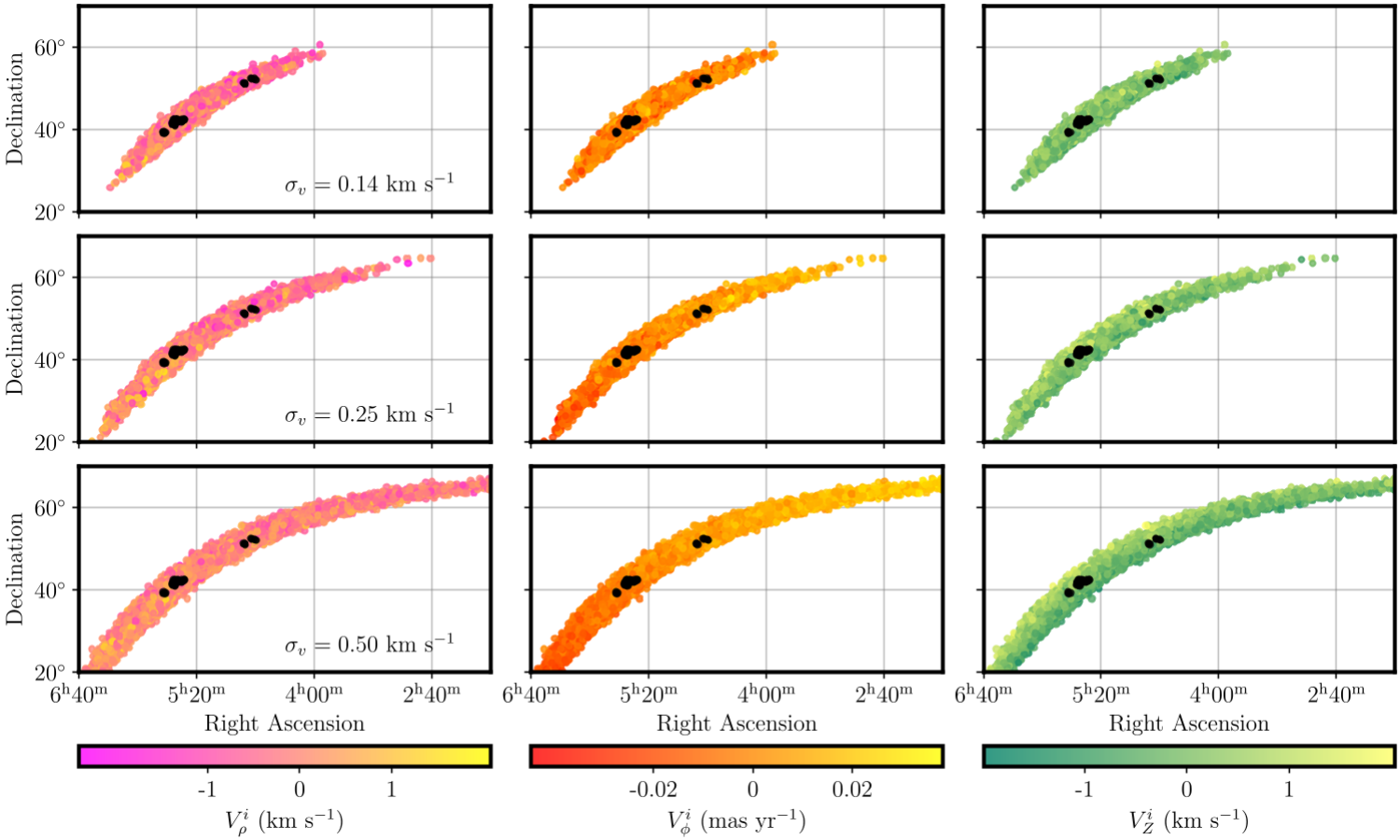}
    \caption{Mock stars integrated forward, color-coded by their initial positions (top 3 rows) and velocities (bottom 3 rows) in Galactocentric cylindrical coordinates, with dynamical members over-plotted as black points. Rows are sorted by the Maxwell-Boltzmann initialization velocity $\sigma_v$. Positions are measured relative to the Galactic center, and velocities are measured relative to the bulk cluster velocity. }
    \label{fig:RA_DEC_birth}
    \end{center}
\end{figure*}

\subsection{Future Dispersion of Theia 456}\label{sec:Future Dispersion of Theia 456}

Ever increasing Gaia precision allows us to detect more and more disperse string-like stellar structures across the sky, which poses the question in relation to Theia 456: at what point will this structure become so dispersed in phase space that it is undetectable as a stellar structure of common origin? Figure \ref{fig:future dispersion} shows model streams initialized at different ages (all with $\sigma_v = 0.14\; \rm km~s^{-1}$, $R = 6$ pc) integrated forward in time; their dispersions in position, proper motion, and parallax increase greatly with age. Even at large ages, the proper motion dispersion is relatively small, though an emergent `tail' appears to branch away from the bulk distribution. Most notably, the dispersion in proper motion for a 350 Myr old model Theia 456 still falls below the threshold adopted in \cite{Cantat-Gaudin_Anders} of: 

\begin{equation}
    \begin{array}{c}
    \sigma_{pm} = \sqrt{\sigma_{\mu_{\alpha^{\star}}}^2 + \sigma_{\mu_{\delta}}^2} \leq 5 \sqrt{2} \frac{\varpi}{4.74 }, \; \varpi > 1 \; \rm{mas }
    \end{array}
\end{equation}
However, the spread across the sky is so large, $\sim 100^{\circ}$ in right ascension and $\sim 80^{\circ}$ in declination, that the field of view in position space for one's chosen clustering method or algorithm would be impractical and overwhelmingly filled with contaminants. We conclude that a stream like Theia 456 will likely be undetectable given our current cluster finding methods in $\lesssim 100$ Myr. One should be careful when generalizing these predictions to other stream-like objects, as their diffusion in phase space due to tidal shredding is greatly affected by their initial velocity distribution, age, and size, as seen in Figures \ref{fig:forward_model_MB_vels} and \ref{fig:forward_model_MB_ages_sizes}. 

\begin{figure*}
    \begin{center}
    \includegraphics[width=1.0\textwidth]{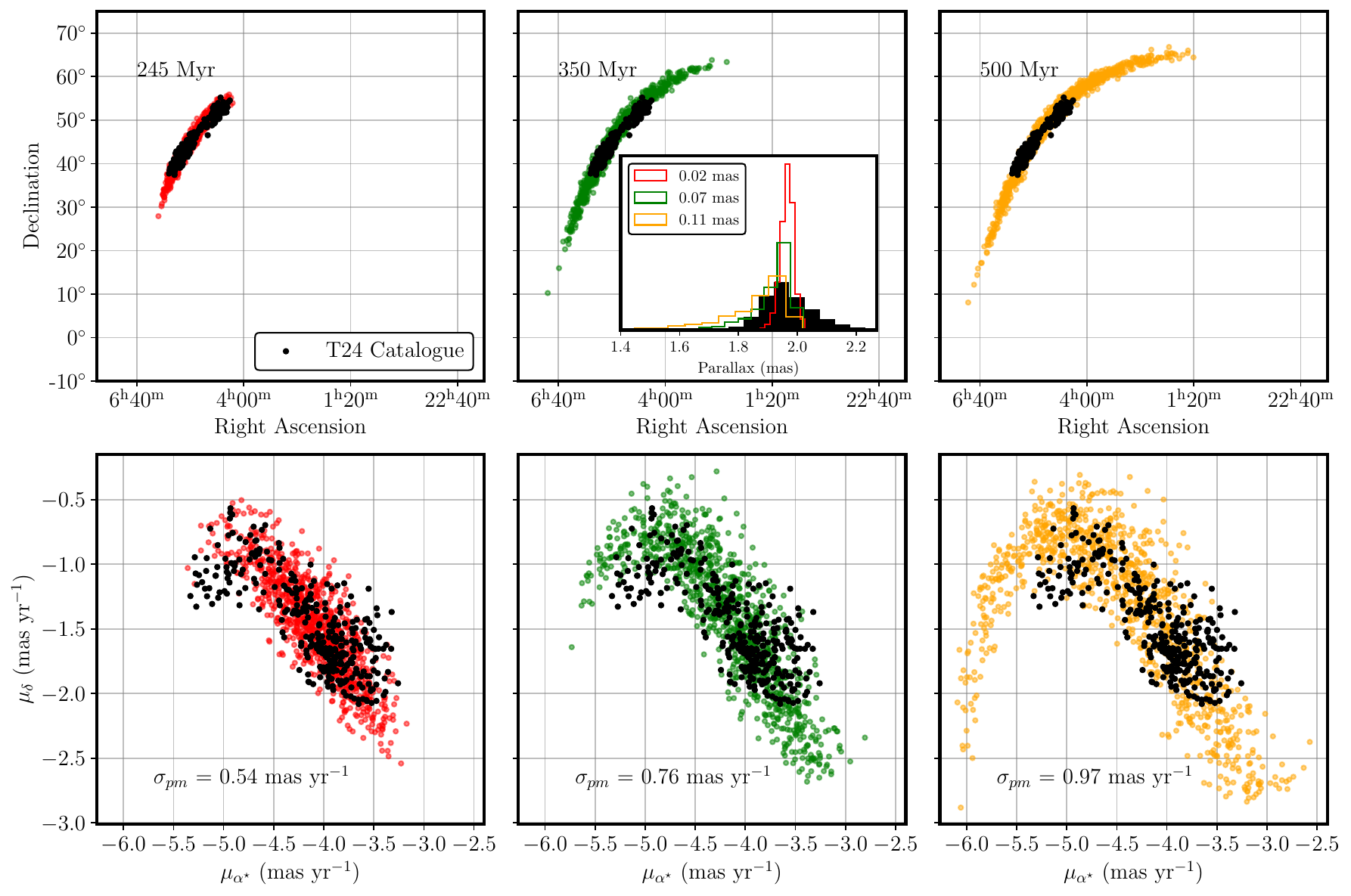}
        \caption{Mock stars (shown as colored points) with different hypothetical ages, denoted in the top left corner of the top panels, forward-integrated, and plotted under our catalog in position and proper motion space. The proper motion dispersion, defined as $\sigma_{pm} = \sqrt{\sigma_{\mu_{\alpha^{\star}}}^2 + \sigma_{\mu_{\delta}}^2}$, is shown for each model stream. The inset shows the density distribution of parallax, and its dispersion in the legend, for each model stream; compared with the entire Theia 456 catalog in solid black.}
    \label{fig:future dispersion}
    \end{center}
\end{figure*}

The distinct tail emerging in proper motion space shown in Figure \ref{fig:future dispersion} points to limitations within our application of clustering algorithms, like \texttt{HDBSCAN}, to blindly classify disperse stellar structures. Currently, algorithms agnostically search for phase space over-densities, and do not consider any physically motivated phase space morphologies. Application of the dynamical modelling outlined in this work could be used to refine catalogs generated by data-driven clustering algorithms. Our results imply that more robust algorithms, trained on physically generated stellar streams in the process of tidal disruption, could provide significant improvement over the current generation of stream-finding algorithms by directly searching for physically motivated phase space over-densities.

\section{Conclusions}\label{sec:conclusions}

In this work, we combine Gaia astrometric quantities with MMT-Hectochelle-derived radial velocities to present a detailed dynamical history of Theia 456, a stellar stream located in the thin disk of the Milky Way. We obtain observations in six fields from MMT along Theia 456's extent and derive precise radial velocities for stars within these fields. We produce a catalog of 321 stars in Theia 456: 43 with full 6D phase space information and the remaining 278 with 5D Gaia astrometry. We quantitatively back-track Theia 456 stars with precise radial velocities under a model Milky Potential and find that its two distinct lobes overlap $\sim 250$ Myr in the past. 

To quantify Theia 456's dynamical evolution, we formulate a Bayesian model, generate an idealized birth distribution of Theia 456 stars, integrate this distribution forward under a model Milky Way potential, and compute the posterior distribution for all 321 stars in Theia 456. We find that Theia 456's origin is well described as a spherically symmetric ball of stars of age $245 \pm 3$ Myr, a scale of $6 \pm 1$ pc (half-mass radius of $9 \pm 2$ pc), and a one-dimensional velocity dispersion of $0.14 \pm 0.02 \; \rm km~s^{-1}$, where all quoted uncertainties are statistical. Our age estimate in particular is highly precise ($\sim 1\%$); typical isochrone uncertainties are on the order of $10\%$ \citep[see][]{Soderblom_2010}, while gyrochronal ages for individual solar type $\sim 200$ Myr old stars are $\sim 38\%$ but may be as good as a few per cent for star clusters \citep{Bouma_2023}. Application of our dynamical model to a variety of clusters may provide more precise age estimates than previously derived, and act as a way to better constrain models of stellar evolution. Additionally, deriving a size and velocity dispersion for a number of post-embedded clusters will provide insight into star-forming regions themselves. 

In our model, we find that parallax uncertainty is the limiting factor in our statistical model, which should improve with Gaia DR4; this will aid in refining cluster memberships, provide a tighter parallax constraint in our dynamical model, and enable radial velocity measurements down to fainter magnitudes. From DR2 to DR3, parallax uncertainties improved by a factor of $\sim 2$ to $5$ (see Figure 2 of \cite{Gaia_DR3_release}); similar improvement in DR4 will greatly improve the reliability of our model. Additional data releases from spectroscopic surveys, like LAMOST and APOGEE, will also aid in providing radial velocities for more and more stars. 

We also vary model parameters in our assumed Milky Way potential and find that reasonable variations in the potential contribute an uncertainty of $5\%$ in age, $50\%$ in radius, and $64\%$ in velocity dispersion. These uncertainties, especially those on the age, are considerably smaller than those found in A22 via isochrone modeling and gyrochronology methods, indicating that dynamical age dating is a very powerful tool, especially for disperse common-origin structures with high quality kinematics at both ends.
    
We then take our generative model and forward integrate those stars to qualitatively assess our model's goodness of fit. A simple spherical model for Theia 456 at birth can reproduce the size and shape of the distribution seen in phase space today. It does not reproduce some of Theia 456's substructure, such as the disperse bridge seen in position space, and the two seperate lobes at the ends of the structure. Reproducing such features may be an observational tool to probe Milky Way disk or initial cluster substructure. Additional follow-up studies of simulated clusters orbiting the Milky Way that include initial substructures, as well as encounters with GMCs and the Milky Way spiral arms, are necessary to disentangle the potential contribution of disk and initial substructure to observed substructure today. 

We also derive an order of magnitude estimate of the initial stellar mass and number of Theia 456. By assuming a Kroupa initial mass function,  we find $m_{\rm total} \approx 900$ M$_{\odot}$ and $n_{\rm total} \approx 2100$ stars. At our derived birth scale, Theia 456 formed as a low density open cluster ($\approx 0.2 \rm ~ M_{\odot}~pc^{-3}$). We do not expect, especially at later times in its evolution, for gravitational interactions between stars in the cluster to play a significant role in the bulk evolution of the cluster, though they may be dynamically important for individual stars.
    
In Figure \ref{fig:RA_DEC_birth}, we show that given a sufficiently low thermal velocity, as found for Theia 456, the location of stars at birth relative to the Galactic center may be well correlated with their position at later times. This may present a way to probe chemical inhomogeneity within molecular clouds, and also mass segregation. This strategy may be the best method of quantifying inhomogeneity within molecular clouds. 
    
We also integrate model stars forward in time from the present day, and find that the Milky Way tides are able to shred these structures quite quickly. Within another $\sim 100$ Myr, the extent of Theia 456 should roughly double in right ascension, making this structure unlikely to be identified by the current generation of clustering algorithms.

Dynamical modelling may also be a powerful tool to refine catalogs produced by data-driven, physics-blind clustering algorithms. It may also be instrumental in pushing the categorization of stellar streams forward by generating new algorithms, trained on generated stellar streams, that search for physically motivated phase space over-densities. Such algorithms may be instrumental in finding additional disperse structures, like Theia 456, that drive our knowledge of the Milky Way's dynamical evolution and star formation history further forward. 

\section*{Acknowledgement}

We thank the anonymous referee for their suggestions that improved the quality of the manuscript. 
The authors thank Carl Rodriguez and David Hogg for useful conversations.
J.C. acknowledges support from the Agencia Nacional de Investigación y Desarrollo (ANID), via Proyecto Fondecyt Regular 1231345, and from ANID
BASAL project CATA2-FB210003. 
Observations reported here were obtained at the MMT Observatory, a joint facility of the Smithsonian Institution and the University of Arizona. This paper uses data products produced by the CfA OIR Telescope Data Center, supported by the Smithsonian Astrophysical Observatory. 

\software{\texttt{AstroPy} \citep{astropy_v3}, \texttt{gala} \citep{gala}, \texttt{MINESweeper} \citep{Cargile2020}, \texttt{NumPy} \citep{Numpy}, \texttt{Matplotlib} \citep{Matplotlib}, \texttt{SciPy} \citep{Scipy}, \texttt{Seaborn} \citep{Seaborn}, \texttt{Pandas} \citep{Pandas}, \texttt{Emcee} \citep{emcee_ref_2013}, \texttt{Corner} \citep{corner}, \texttt{IMF} (\url{https://github.com/keflavich/imf}), \texttt{pytreegrav} (\url{https://github.com/mikegrudic/pytreegrav})} 

\appendix{}

\begin{deluxetable}{ccccccccccccccc}[h!]
\rotate
 \tablecaption{Theia 456 Catalog \label{tab:catalog}}
 
 \tablehead{
    \colhead{Gaia DR3 ID} & 
    \colhead{$S/N$} & 
    \colhead{$\frac{V_{\rm rad}}{\rm km~s^{-1}}$} &
    \colhead{$\frac{V_{\rm rot}}{\rm km~s^{-1}}$} &
    \colhead{$\frac{\rm Dist.}{\rm pc}$} & 
    \colhead{$\frac{\varpi}{\rm mas}$} &
    \colhead{$\frac{A_V}{\rm mag}$} & 
    \colhead{$\frac{T_{\rm eff}}{\rm K}$} & 
    \colhead{$\frac{M}{\rm \Msun}$} &
    \colhead{$\frac{\rm [Fe/H]}{\rm dex}$} & 
    \colhead{$\frac{\rm [\alpha/H]}{\rm dex}$} &
    \colhead{$\log\big(\frac{g}{\rm cm \; s^{-2}} \big)$} & 
    \colhead{$\log\big(\frac{R}{\rm \Rsun} \big)$} & 
    \colhead{$\log\big(\frac{\rm Age}{\rm yr} \big)$} & 
    \colhead{$\log\big(\frac{L}{\rm \Lsun} \big)$}  
    }
\startdata \vspace{0.1cm}
194453957134786176 & 114 & $-$13.77 & 47.9 & 500 & 1.98 & 0.36 & 6459 & 1.243 & 0.06 & $-$0.08 & 4.355 & 0.088 & 9.0 & 0.37 \\
194450491096608000 & 18 & $-$11.90 & 4.5 & 580 & 1.72 & 0.57 & 4774 & 0.746 & $-$0.27 & 0.16 & 4.660 & $-$0.144 & 7.8 & $-$0.62 \\
194445062257874048 & 30 & $-$13.62 & 0.4 & 530 & 1.88 & 0.45 & 5063 & 0.849 & $-$0.05 & 0.08 & 4.620 & $-$0.104 & 8.1 & $-$0.44 \\
194400944354088960 & 20 & $-$13.76 & 0.9 & 530 & 1.90 & 0.51 & 4878 & 0.794 & $-$0.13 & 0.11 & 4.632 & $-$0.120 & 8.4 & $-$0.53 \\
194387097379628672 & 32 & $-$13.97 & 7.6 & 500 & 1.99 & 0.56 & 5489 & 0.933 & $-$0.05 & 0.07 & 4.572 & $-$0.074 & 8.7 & $-$0.24 \\
... & ... & ... & ... & ... & ... & ... & ... & ... & ... & ... & ... & ... & ... & ... \\
191391718467413504 & 49 & $-$13.90 & 7.9 & 510 & 1.95 & 0.33 & 5612 & 0.960 & $-$0.01 & 0.02 & 4.566 & $-$0.069 & 8.2 & $-$0.19 \\
191365566412178048 & 138 & $-$13.43 & 22.3 & 500 & 2.00 & 0.32 & 6645 & 1.304 & 0.04 & $-$0.06 & 4.308 & 0.123 & 9.0 & 0.49 \\
191362890644144128 & 10 & $-$14.25 & 4.9 & 510 & 1.97 & 1.09 & 4635 & 0.774 & $-$0.16 & 0.22 & 4.635 & $-$0.121 & 8.7 & $-$0.62 \\
191357324369704448 & 32 & $-$13.31 & 0.3 & 530 & 1.90 & 0.35 & 5211 & 0.867 & $-$0.07 & 0.05 & 4.607 & $-$0.098 & 8.5 & $-$0.38 \\
190222319132712832 & 21 & $-$13.05 & 0.6 & 550 & 1.80 & 0.78 & 4952 & 0.811 & $-$0.10 & 0.08 & 4.640 & $-$0.121 & 7.8 & $-$0.51 \\
\enddata
\tablenotetext{}{An abbreviated data table for our catalog of stars in Theia 456. Stars listed above come from the subset of stars with MMT-Hectochelle spectra and derived stellar parameters. A machine readable format is available for download.}

\end{deluxetable}

\clearpage

\bibliography{references}{}
\bibliographystyle{aasjournal}

\end{document}